\documentclass[11pt,preprintnumbers,superscriptaddress,nofootinbib]{revtex4}

\usepackage{url}
\usepackage{graphicx}
\usepackage{dcolumn}
\usepackage{bm}
\usepackage{amssymb}
\usepackage{amsmath}
\usepackage{epsfig}    
\usepackage{color}
\usepackage{slashed}
\usepackage{hyperref}
\usepackage{comment}
\usepackage[utf8]{inputenc}
\allowdisplaybreaks

\newcommand{\sba}{s_{\beta-\alpha}}
\newcommand{\cba}{c_{\beta-\alpha}}

\newcommand{\lam}{\lambda}
\newcommand{\sig}{\sigma}
\newcommand{\eeha}{e^+e^-\to h\gamma}
\newcommand{\haa}{h\to\gamma\gamma}
\newcommand{\hza}{h\to Z\gamma}
\newcommand{\drha}{\Delta R_{h\gamma}}
\newcommand{\draa}{\Delta R_{\gamma\gamma}}
\newcommand{\drza}{\Delta R_{Z\gamma}}
\newcommand{\mc}{m_{H^+}}
\newcommand{\mcc}{m_{H^{++}}}

\begin{document} 


\title{Single Higgs production in association with a photon\\
at electron--positron colliders in extended Higgs models
}

\author{Shinya Kanemura}
\affiliation{Department of Physics, Osaka University, Toyonaka, Osaka 560-0043, Japan}

\author{Kentarou Mawatari}
\affiliation{Department of Physics, Osaka University, Toyonaka, Osaka 560-0043, Japan}

\author{Kodai Sakurai}
\affiliation{Department of Physics, University of Toyama, 3190 Gofuku, Toyama 930-8555, Japan}
\affiliation{Department of Physics, Osaka University, Toyonaka, Osaka 560-0043, Japan}

\preprint{OU-HET 976}
\preprint{UT-HET 126}

\begin{abstract}
\vspace*{5mm}
\begin{center}
{\bf Abstract}
\end{center}
We study associated Higgs production with a photon at electron--positron colliders, $e^+e^-\to h\gamma$, in various extended Higgs models, such as the inert doublet model (IDM), the inert triplet model (ITM)   and the two Higgs doublet model (THDM).
The cross section in the standard model (SM) is maximal around $\sqrt{s}=250$~GeV, and we present how and how much the new physics can enhance or reduce the production rate.
We also discuss the correlation with the $h\to\gamma\gamma$ and $h\to Z\gamma$ decay rates.
We find that, with a sizable coupling to a SM-like Higgs boson, charged scalars can give considerable contributions to both the production and the decay if their masses are around 100~GeV.
Under the theoretical constraints from vacuum stability and perturbative unitarity as well as the current constraints from the Higgs measurements at the LHC, the production rate can be enhanced from the SM prediction at most by a factor of two in the IDM. 
In the ITM, in addition, we find a particular parameter region where the $h\gamma$ production significantly increases by a factor of about six to eight, 
but the $h\to\gamma\gamma$ decay still remains as in the SM.
In the THDM, possible deviations from the SM prediction are minor in the viable parameter space.
\end{abstract}

\maketitle


\newpage
\section{Introduction}\label{sec:intro}

High-energy $e^+e^-$ colliders as a Higgs factory have been discussed for a long time
in order to identify the Higgs sector; i.e., a mechanism of electroweak symmetry breaking
and its relation to physics beyond the standard model (SM).
Especially after the discovery of the 125~GeV Higgs boson in $pp$ collisions at the Large Hadron Collider (LHC),
the physics potential for precise measurements of the Higgs couplings 
at $\sqrt{s}=240-250$~GeV has been extensively studied for realization of the International Linear Collider (ILC)~\cite{Baer:2013cma,Fujii:2017vwa,Asai:2017pwp} as well as 
the lepton collision option of the Future Circular Collider (FCC-ee)~\cite{Gomez-Ceballos:2013zzn} and  
the Circular Electron Positron Collider (CEPC)~\cite{CEPC-SPPCStudyGroup:2015csa}.
Such collision energies are optimal for studying associated Higgs production with a $Z$ boson, $e^+e^-\to hZ$, 
where the total cross section is maximal around $\sqrt{s}=250$~GeV in the SM.

On the other hand, it is less known that with the mass of $m_h=125$~GeV the cross section for associated Higgs production with a photon, 
$e^+e^-\to h\gamma$, also has a peak around $\sqrt{s}=250$~GeV in the SM~\cite{Abbasabadi:1995rc,Djouadi:1996ws}.
Since the process is protected by the electromagnetic gauge symmetry and 
the tree-level contribution is highly suppressed by the electron mass, it is essentially loop induced, and hence
the cross section is rather small, ${\cal O}(0.1)$~fb at $\sqrt{s}=250$~GeV.
However, because the signal is very clean; i.e., a monochromatic photon in the final state, 
the above future colliders with the design luminosity may be able to observe the signal.
Moreover, new physics can substantially enhance the production rate relative to the SM case;
i.e., the process has the potential to explore physics beyond the SM.
In this work, we study how and how much new physics can enhance (or reduce) the $h\gamma$ signal at future $e^+e^-$ colliders. 

The process was studied in the SM long time ago~\cite{Barroso:1985et,Abbasabadi:1995rc,Djouadi:1996ws}.
The next-to-leading-order QCD corrections were recently estimated in Ref.~\cite{Sang:2017vph},
which shows negligible impact at lower center-of-mass energies $\sqrt{s}<300$~GeV.
On the other hand, new physics effects on the process have not been fully explored yet.  
There are several studies in the literature in the context of anomalous Higgs-boson couplings or an effective field theory~\cite{Gounaris:1995mx,Mahanta:1997hg,Ren:2015uka,Cao:2015iua,Li:2015kxc,Kobakhidze:2016mfx}
as well as concrete new physics models such as supersymmetric models~\cite{Djouadi:1996ws,Hu:2014eia,Gounaris:2015tna,Heinemeyer:2015qbu}
and the inert doublet model (IDM)~\cite{Arhrib:2014pva}.

In this article, we 
study the $e^+e^-\to h\gamma$ process
in various extended Higgs models, including the IDM studied in Ref.~\cite{Arhrib:2014pva}, the inert triplet model (ITM) and two Higgs doublet models (THDMs) as distinctive benchmark models for the process.
We calculate the loop-induced amplitudes at the leading order (LO) by
employing and extending the {\sc H-COUP} program~\cite{Kanemura:2017gbi}, which evaluates the renormalized gauge-invariant vertex functions for the SM-like Higgs boson in various extended Higgs models. 
Through a systematic study of the different models, we would like to obtain more general information on new physics effects in the signal. 

We begin with a brief review of the $e^+e^-\to h\gamma$ cross section in the SM in Sec.~\ref{sec:sm}.
After introducing each extended Higgs model in Sec.~\ref{sec:model}, 
we present the cross sections in each model, and also discuss the correlations with the $\haa$ and $\hza$ decay rates in Sec.~\ref{sec:results}. 
The results are shown not only at $\sqrt{s}=250$~GeV but also at $\sqrt{s}=500$~GeV
in anticipation of the energy extension of the ILC or the Compact Linear Collider (CLIC)~\cite{CLIC:2016zwp}.
The summary is given in Sec.~\ref{sec:summary}.
Explicit formulas for the $h\gamma\gamma$ and $hZ\gamma$ vertex functions are listed in the Appendix.

\section{Process of $e^+e^-\to h\gamma$ in the SM}\label{sec:sm}

\begin{figure}
 \center 
 \includegraphics[width=0.5\textwidth]{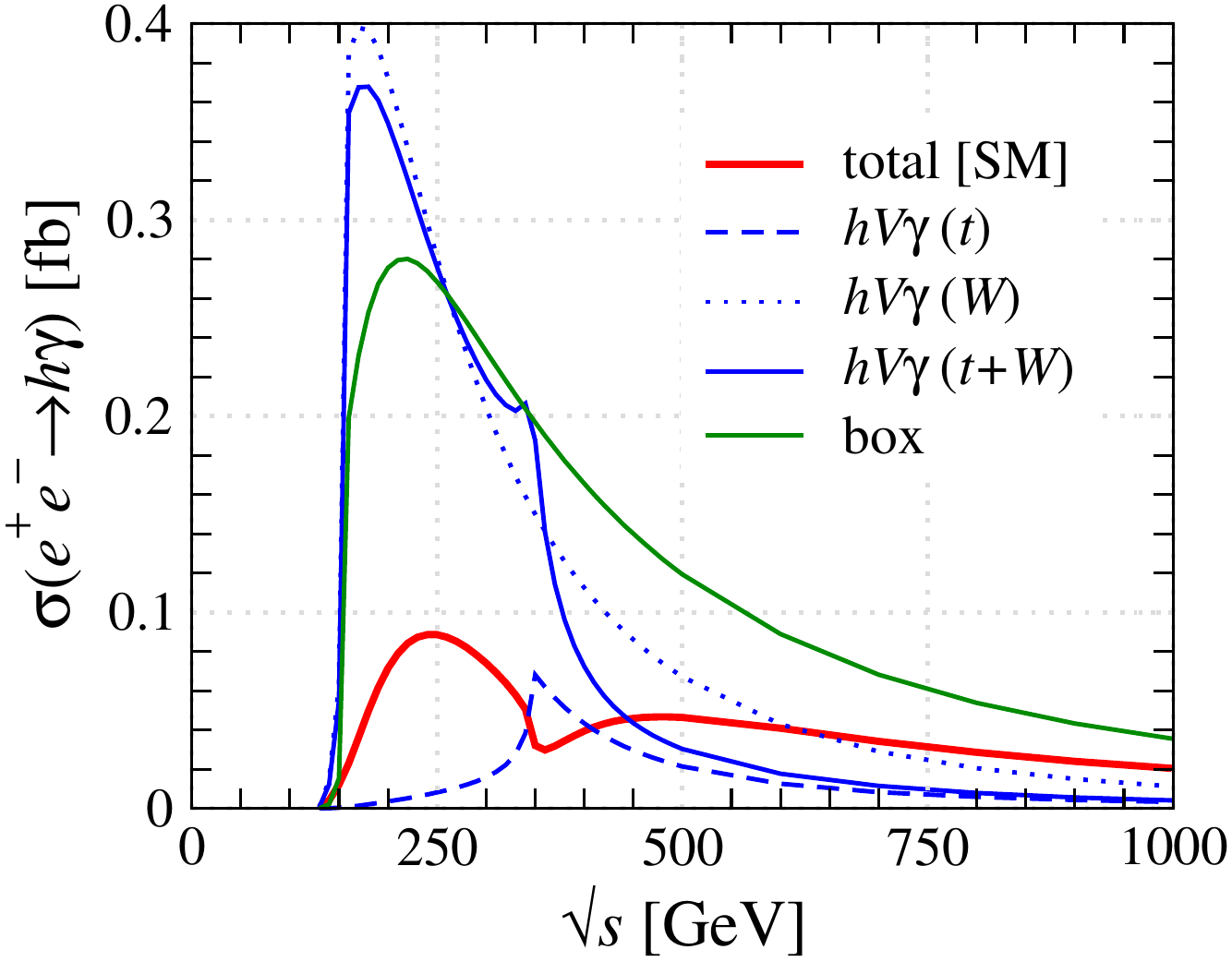}
 \caption{Total cross section for $e^+e^-\to h\gamma$ in the SM as a function of the collision energy.
 The total rate (red solid) is decomposed into two contributions;
 triangle loops (blue solid) and box loops (green solid).
 The contribution from the triangle loops is further decomposed into the ones from top quarks (blue dashed) and from $W$ bosons (blue dotted).
 }
\label{fig:xsec_sm}
\end{figure}

We briefly review the total cross section for $e^+e^-\to h\gamma$ in the SM~\cite{Barroso:1985et,Abbasabadi:1995rc,Djouadi:1996ws,Sang:2017vph}.
As mentioned in Sec.~\ref{sec:intro}, the process is loop induced, and we evaluate it at LO.
The details of the calculation will be discussed in Sec.~\ref{sec:results}.

\begin{figure}
 \center 
 \includegraphics[width=1\textwidth]{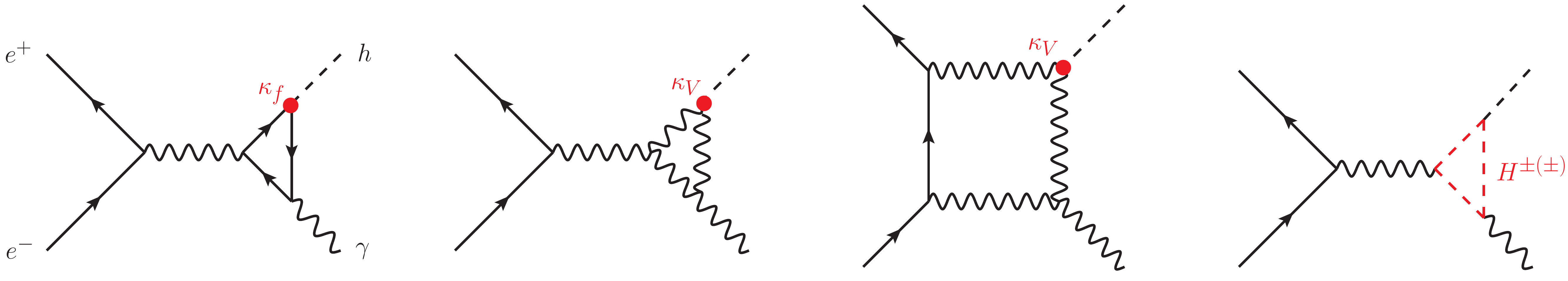}\\[-2mm] 
 \hspace*{1mm}(a)\hspace*{37mm}(b)\hspace*{37mm}(c)\hspace*{37mm}(d) 
 \caption{Representative Feynman diagrams for $e^+e^-\to h\gamma$.}
\label{fig:diagram}
\end{figure}

Figure~\ref{fig:xsec_sm} shows the total cross section $\sigma(e^+e^-\to h\gamma)$ in the SM
as a function of the collision energy $\sqrt{s}$.%
\footnote{The cross section is shown with unpolarized beams. With fully polarized left-handed electrons and right-handed positrons, the cross section is approximately four times larger than in the unpolarized case~\cite{Djouadi:1996ws}.} 
One can observe
 an interesting peak-dip structure due to negative interference 
among different contributions~\cite{Djouadi:1996ws,Sang:2017vph}.
To understand the structure, we decompose all the contributions into two categories in the 't Hooft--Feynman gauge; 
one is contributions from triangle $hV\gamma$ ($V=\gamma$ or $Z$) loops, and the other is from box loops, 
shown in blue and green lines, respectively, in Fig.~\ref{fig:xsec_sm}.
The $hV\gamma$ contributions are further decomposed into the ones from top quarks and from $W$ bosons, denoted by blue dashed and dotted lines, respectively.
We depict the representative Feynman diagrams in Fig.~\ref{fig:diagram}. 
The contribution from fermion loops (diagram (a)), mainly from top quarks, becomes maximum around the $t\bar t$ threshold, leading to the dip structure for the total rate. 
$W$-boson loops in the triangle loop contribution (diagram (b)) are dominant over the top-quark loops, especially $\sqrt{s}<2m_t$.
The box contribution includes $W/\nu$ loops (diagram (c)) as well as a $Z/e$ loop. 
Combining all the contributions at the amplitude level, the total cross section eventually has
the first peak at $\sqrt{s}\sim 250$~GeV and the second one at $\sqrt{s}\sim 500$~GeV.
As we see, the production rate is very sensitive to the magnitudes of each amplitude as well as the relative phases among them. 
Therefore, slight modifications of the SM interactions as well as loop contributions from new particles
can substantially enhance (or reduce) the production rate. 
We note that the relative magnitudes between the triangle and box contributions are rather different 
at different energies, and hence new physics can affect the total rates differently at different collision energies, e.g., between at $\sqrt{s}=250$ and 500~GeV, as seen below.  

Before turning to the next section, we discuss how new physics can affect the process.
In extended Higgs models, the SM-like Higgs interactions can be modified due to the mixing with other neutral scalars.
The effect is described as the tree-level scaling factors 
\begin{align}
  \kappa_X=\frac{g_{hXX}^{\rm NP}}{g_{hXX}^{\rm SM}},
\label{kappa}
\end{align}
the ratio of the Higgs coupling in a new physics model to the one in the SM, denoted as red blobs in
diagrams (a), (b) and (c) in Fig.~\ref{fig:diagram}.
In addition, the process can be modified by additional triangle loops of charged scalars; 
see diagram (d).
In the following, we study such effects in each extended Higgs model.

\section{Extended Higgs models}\label{sec:model}

As discussed above, in extended Higgs models, the $e^+e^-\to h\gamma$ process can be altered through 1) the modified couplings of the SM-like Higgs boson and/or
2) charged scalar loops. 
In order to study such new physics effects on the process,
we consider three distinctive models: the inert doublet model (IDM), the inert triplet model (ITM) with the hypercharge $Y=1$, and the two Higgs doublet model (THDM). 
At the tree level, while
in the first two models the SM-like Higgs couplings to SM fermions and weak bosons are the same as in the SM, in the THDM those are modified by the mixing of the scalar fields. 
While all the three models include singly charged scalars, the ITM includes doubly charged scalars as well.
We briefly describe the three models in order.
We give the Higgs potential and define the relevant parameters for each model.
We also describe theoretical and experimental constraints on the parameters.
The experimental constraint that is strongly relevant to the $\eeha$ process comes from the $h\to\gamma\gamma$ signal at the LHC,
which will be discussed exceptionally in the next section.

\subsection{Inert doublet model}\label{sec:idm}

In the IDM~\cite{Deshpande:1977rw,Barbieri:2006dq}, an isospin doublet field $\Phi_2$ with hypercharge $Y=1/2$ is added to 
the SM Higgs doublet field $\Phi_1$.
The model imposes an exact $Z_2$ symmetry under which $\Phi_2$ is odd while all the other fields are even.
The general Higgs potential under the $Z_2$ symmetry is given by
\begin{align}
V(\Phi_1,\Phi_2) &= \mu_1^2|\Phi_1|^2+\mu_2^2|\Phi_2|^2 \notag\\
&\quad+\frac{1}{2}\lambda_1|\Phi_1|^4+\frac{1}{2}\lambda_2|\Phi_2|^4+\lambda_3|\Phi_1|^2|\Phi_2|^2+\lambda_4|\Phi_1^\dagger\Phi_2^{}|^2
+\frac{1}{2}\lambda_5\{(\Phi_1^\dagger\Phi_2^{})^2+\text{h.c.}\},
\label{pot_idm}
\end{align}
where all the parameters can be taken to be real.
Because of the $Z_2$ symmetry, only $\Phi_1$ can have a nonzero vacuum expectation value (VEV) $v$.
The two doublet fields $\Phi_{1,2}$ are parameterized as 
\begin{align}
\Phi_1=\left(\begin{array}{c}
G^+\\
\frac{1}{\sqrt{2}}(v+h+iG^0)
\end{array}\right),\quad
\Phi_2=\left(\begin{array}{c}
H^+\\
\frac{1}{\sqrt{2}}(H+iA)
\end{array}\right),
\label{idm_f}
\end{align}
where $G^{\pm,0}$ are Nambu--Goldstone bosons to be absorbed by weak bosons.
$h$ is the SM-like Higgs boson with the mass of 125~GeV, while $H$, $A$ and $H^{\pm}$ are $Z_2$-odd inert scalar bosons.
After imposing the stationary condition, masses of the scalar bosons are written in terms of the parameters in the potential as 
\begin{align}
 m_h^2=-2\mu_1^2=v^2\lambda_1,\quad
 m_{H^\pm}^2=\mu_2^2+\frac{1}{2}v^2\lambda_3,\quad
 m_H^2=\mu_2^2+\frac{1}{2}v^2\lambda_{345}^+,\quad
 m_A^2=\mu_2^2+\frac{1}{2}v^2\lambda_{345}^-,
\label{mass_idm}
\end{align}
with $\lambda_{345}^{\pm}\equiv\lambda_3+\lambda_4\pm\lambda_5$.
We choose the following five parameters in the IDM:%
\footnote{Instead of $\lambda_3$, the H-COUP program uses $\mu_2^2$ as an input parameter.}
\begin{align}
 m_H,~m_A,~m_{H^+},~\lambda_2,~\lambda_3.
\end{align}

Because of the exact $Z_2$ symmetry, none of the inert scalars has direct couplings to SM fermions.
The couplings of the SM-like Higgs boson to SM fermions and weak bosons are exactly the same as in the SM;
i.e.,
\begin{align}
 \kappa_f=\kappa_V=1
\label{kappa_idm}
\end{align}  
in Eq.~\eqref{kappa}.%
\footnote{Note that those couplings deviate from the SM prediction 
at higher orders~\cite{Arhrib:2015hoa,Kanemura:2016sos}.}
While the couplings of the charged scalars to gauge bosons are the SM gauge couplings,
the coupling to the SM-like Higgs boson is given by
\begin{align}
 g_{hH^+H^-} =-\frac{2}{v}(m_{H^+}^2-\mu_2^2) =-v\lambda_3.
\label{ghHH_idm}
\end{align}
We note that, if $\lambda_3=0$ (or $m_{H^+}^2=\mu_2^2$), the predictions for $e^+e^-\to h\gamma$ in the IDM and in the SM are identical at the LO.

The following theoretical and experimental constraints on the above parameters are taken into account~\cite{Swiezewska:2012ej,Swiezewska:2012eh}.
Vacuum stability~\cite{Deshpande:1977rw} and the existence of the inert vacuum~\cite{Ginzburg:2010wa} require
\begin{align}
 \lam_{1,2}>0,\quad
 \sqrt{\lam_1\lam_2}+\lam_3+{\rm MIN}[0,\lam_4-\lam_5,\lam_4+\lam_5]>0,
\end{align} 
and
\begin{align}
 m_{h,H,A,H^{+}}^2>0,\quad
 \frac{\mu_1^2}{\sqrt{\lam_1}}<\frac{\mu_2^2}{\sqrt{\lam_2}},
\end{align} 
respectively.
We evaluate constraints from perturbative unitarity, adopting the formulas given in Refs.~\cite{Kanemura:1993hm,Akeroyd:2000wc}.%
\footnote{
Notice that if we use criteria other than the tree-level unitarity such as
the unitarity bounds at one-loop level~\cite{Grinstein:2015rtl} 
or the triviality bounds~\cite{Kanemura:1999xf,Barbieri:2006dq,Goudelis:2013uca}, 
the parameter space can be more restricted.
}

LEP-I precision measurements exclude the possibility that weak bosons decay into a pair of inert scalars; i.e.,
$m_{H^+}+m_{H,A}>m_W,\ 2m_{H^+}>m_Z,\ {\rm and}\ m_{H}+m_{A}>m_Z$.
Moreover, the electroweak precision data, especially the $T$ parameter, imply that 
a mass difference between charged scalars ($H^\pm$) and a neutral inert scalar ($H$ or $A$) must not be too large~\cite{Barbieri:2006dq,Belyaev:2016lok}.
We note that direct searches for additional scalars in collider experiments have been often done in the context of THDMs or supersymmetric models, in which the scalar particles couple to SM fermions.
Therefore, careful reinterpretations are necessary to constrain the inert models.
LEP-II chargino search results are naively interpreted as $m_{H^+}>70-90$~GeV~\cite{Pierce:2007ut},
while neutralino search results exclude the region where simultaneously 
$m_{H}<80$~GeV, $m_{A}<100$~GeV, and $m_{A}-m_{H}>8$~GeV,
where $m_H<m_A$ is assumed~\cite{Lundstrom:2008ai,Gustafsson:2010zz}.  
The constraints from the 8~TeV LHC data have been also studied recently~\cite{Belanger:2015kga,Belyaev:2016lok}. 

\begin{figure}
 \includegraphics[height=0.5\textwidth]{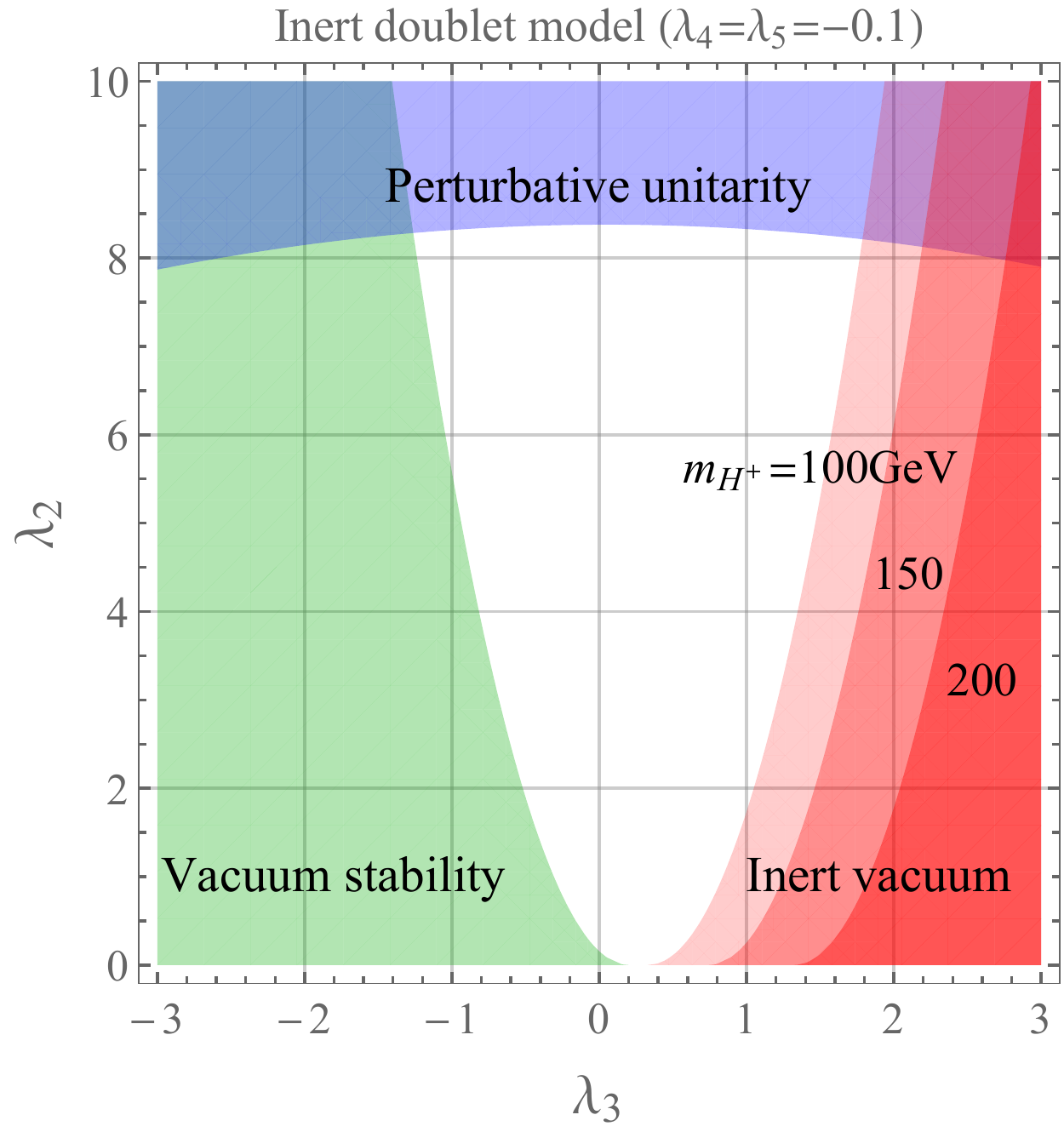}\qquad
 \includegraphics[height=0.5\textwidth]{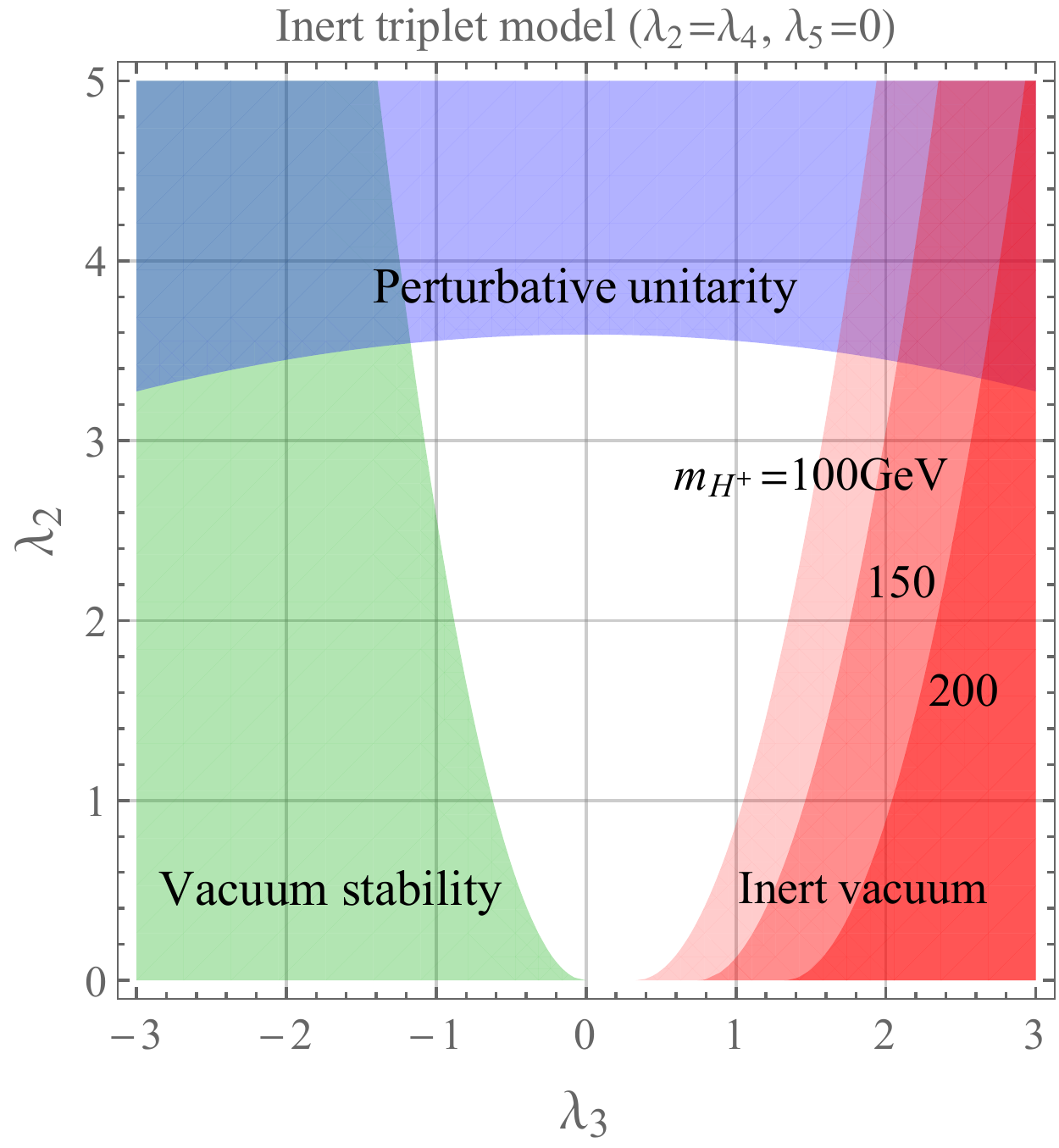}
 \caption{Theoretical constraints on the $\lam_3$--$\lam_2$ plane in the IDM (left) and in the ITM (right).}
\label{fig:thconst}
\end{figure}

In our study, the value of the coupling $g_{hH^+H^-}$ in Eq.~\eqref{ghHH_idm}; i.e., $\lam_3$, 
is one of the relevant parameters. 
Figure~\ref{fig:thconst}(left) shows allowed regions of $\lam_3$ as a function of $\lam_2$ in the IDM
after imposing the above theoretical constraints.
Here, we take $\lam_4=\lam_5=-0.1$ as a benchmark to give a mass spectrum as $m_H<\mc=m_A$.%
\footnote{See Ref.~\cite{Swiezewska:2012ej} for the allowed region in detail.}
$\lam_3$ is bounded below by the vacuum stability (green region) and 
above by the existence of the inert vacuum (red).
The constraint from the inert vacuum depends on the charged scalar mass.
Those constrains depend on $\lam_2$, which is bounded above by perturbative unitarity (blue).

We note that the IDM has been often studied in the context of dark matter~\cite{Ma:2006km,Barbieri:2006dq,LopezHonorez:2006gr} 
since the lightest neutral inert scalar can be stable due to the discrete $Z_2$ symmetry.
Although we do not consider such a property in this work, 
dark matter constraints, 
such as from relic density and direct detection, can also limit the parameter space;
see, e.g., Ref.~\cite{Belyaev:2016lok} for details.%
\footnote{Especially, the recent result of the XENON-1T dark matter direct detection~\cite{Aprile:2017iyp}
provides a strong limit on the mass of the dark matter candidate; i.e., the lightest inert scalar, as well as $\lambda_{345}$~\cite{Ilnicka:2018def}.
We note that, as mentioned below in footnote 9, we can still find interesting parameter regions which are compatible with dark matter constraints.
}

\subsection{Inert triplet model}\label{sec:itm}

Instead of a doublet field as in the IDM, 
we introduce an isospin triplet field $\Delta$ with hypercharge $Y=1$,
and impose an exact $Z_2$ symmetry under which $\Delta$ is odd while all the other SM fields are even~\cite{Hambye:2009pw,Araki:2011hm}.
The Higgs potential is given by%
\footnote{For comparison with the IDM, we adopt the convention in Ref.~\cite{Hambye:2009pw} for the definition of the potential,
which differs from the usual convention in which $\lam_3$ and $\lam_4$ are exchanged.}
\begin{align}
V(\Phi,\Delta) &= \mu_1^2|\Phi|^2+\mu_2^2{\rm Tr}[\Delta^\dagger\Delta] \notag\\
&\quad+\frac{1}{2}\lambda_1|\Phi|^4+\frac{1}{2}\lambda_2({\rm Tr}[\Delta^\dagger\Delta])^2
+\lambda_3|\Phi|^2{\rm Tr}[\Delta^\dagger\Delta]
+\frac{1}{2}\lambda_4{\rm Tr}[(\Delta^\dagger\Delta)^2]
+\lambda_5 \Phi^\dagger\Delta\Delta^\dagger\Phi,
\label{pot_itm}
\end{align}
where all the parameters are real, and the fields are parameterized by
\begin{align}
\Phi=\left(\begin{array}{c}
G^+\\
\frac{1}{\sqrt{2}}(v+h+iG^0)
\end{array}\right),\quad
\Delta=\left(\begin{array}{cc}
\frac{1}{\sqrt{2}}H^+ &H^{++}\\
\frac{1}{\sqrt{2}}(H+iA) & -\frac{1}{\sqrt{2}}H^+
\end{array}\right).
\label{itm_f}
\end{align}
In addition to singly charged scalars, the model contains doubly charged scalars.
The mass spectrum for the inert scalars is given by
\begin{align}
 m_{H^{\pm\pm}}^2=\mu_2^2+\frac{1}{2}v^2\lambda_3,\quad
 m_{H,A}^2=\mu_2^2+\frac{1}{2}v^2(\lambda_3+\lambda_5),\quad
 m_{H^\pm}^2=\frac{1}{2}(m_{H^{\pm\pm}}^2+m_{H}^2),
\label{mass_itm}
\end{align}
while the mass of the SM-like Higgs boson is $m_h^2=-2\mu_1^2=v^2\lambda_1$.
We choose the following five parameters in the ITM:
\begin{align}
 m_H,~m_{H^{++}},~\lam_2,~\lambda_3,~\lam_4.
\end{align}

Similar to the IDM, the inert scalars do not couple to SM fermions, 
and the SM-like Higgs boson interacts with SM fermions and weak bosons as in the SM; i.e.,
\begin{align}
 \kappa_f=\kappa_V=1.  
\label{kappa_itm}
\end{align}  
The couplings of the charged scalars to the SM-like Higgs boson are given by
\begin{align}
 g_{hH^{++}H^{--}} &=-\frac{2}{v}(m_{H^{++}}^2-\mu_2^2) =-v\lambda_3, \label{ghHH2_itm} \\ 
 g_{hH^+H^-} &=-\frac{2}{v}\Big\{\frac{1}{2}(m_{H^{++}}^2+m_{H^{}}^2)-\mu_2^2\Big\}
  =-v\Big(\lambda_3+\frac{1}{2}\lambda_5\Big).  
\label{ghHH_itm}
\end{align}
We note that, if $\lam_5=0$, the doubly and singly charged scalars have the same mass and the same coupling to the SM-like Higgs boson.
If $\lambda_3=\lambda_5=0$; i.e., $m_{H^{++}}^2=m_{H^+}^2=\mu_2^2$, the predictions for $e^+e^-\to h\gamma$ in the ITM and in the SM are identical at LO.

The above parameters should satisfy the following theoretical and experimental constraints. 
With the squared masses in Eq.~\eqref{mass_itm} positive,
vacuum stability~\cite{Arhrib:2011uy} and the existence of the inert vacuum require  
\begin{align}
&\lambda_1 > 0,\quad 
\lambda_2+\text{MIN}[\lambda_4,\lambda_4/2] > 0,\quad \notag\\
&\lambda_3+\text{MIN}[0,\lambda_5] +\text{MIN}\big[\sqrt{\lambda_1(\lambda_2+\lambda_4)},\sqrt{\lambda_1(\lambda_2+\lambda_4/2)}\big]> 0,
\label{vs_condition}
\end{align}
and
\begin{align}
  \frac{\mu_1^2}{\sqrt{\lam_1}}<\frac{\mu_2^2}{\sqrt{\lam_2+\lam_4}},
\end{align}
respectively.
We evaluate constraints from perturbative unitarity adopting the formulas given in Refs.~\cite{Aoki:2007ah,Arhrib:2011uy,Hartling:2014zca}.

The constraints on singly charged scalars from LEP measurements are similar to the ones in the IDM described above~\cite{Chun:2012jw}.
Doubly charged scalars have been searched at the LEP~\cite{Abbiendi:2001cr,Abdallah:2002qj,Achard:2003mv}, Tevatron~\cite{Abazov:2011xx,Aaltonen:2011rta},
and LHC~\cite{Chatrchyan:2012ya,ATLAS:2014kca},
and were excluded up to $\mcc\sim400-550$~GeV.
However, those analyses assume a 100\% branching ratio for the $H^{\pm\pm}$ decay to charged leptons, and hence those limit cannot be applied for the ITM.  
Constraints on the ITM at the LHC were studied in Ref.~\cite{Ayazi:2014tha}
but have not been fully explored yet.
We note that upper limits on the cross section times the branching ratio for doubly charged Higgs boson produced in vector-boson fusion, $\sigma(pp\to H^{\pm\pm}jj)\times B(H^{\pm\pm}\to W^{\pm}W^{\pm})$, was reported recently in Ref.~\cite{Sirunyan:2017ret}.

Similar to the IDM, one of the relevant parameters is $\lam_3$.
In Fig.~\ref{fig:thconst}(right), we show allowed regions of $\lam_3$ as a function of $\lam_2$ in the ITM
after imposing the theoretical constraints from vacuum stability (green region), the existence of the inert vacuum (red), and 
perturbative unitarity (blue), where we take $\lam_2=\lam_4$ and $\lam_5=0$ for simplicity.%
\footnote{See Ref.~\cite{Aoki:2012jj} for the $\lam_5\ne0$ case.}
The allowed region of $\lam_3$ in the ITM is slightly smaller than that in the IDM.

We note that the ITM with hypercharge $Y=0$ does not contain doubly charged scalars, and hence
the situation is very similar to that in the IDM in our study.
Similar to the IDM, the ITM has been studied in the context of dark matter~\cite{Hambye:2009pw,Araki:2011hm,Ayazi:2014tha}.
For the hypercharge $Y=1$ case, however, most of the parameter space is already excluded by the direct detection experiments due to the $Z$-mediated tree-level scattering~\cite{Hambye:2009pw,Araki:2011hm}.%
\footnote{By adding additional particle contents in the ITM, we can avoid such dark matter constraints; 
see, e.g., Refs.~\cite{Kajiyama:2013zla,Okada:2015vwh,Lu:2016dbc}.}
Here, we consider the ITM with $Y=1$ as a benchmark model just to describe new physics effects on the $\eeha$ processes.

\subsection{Two Higgs doublet models}\label{sec:thdm}

The THDM is similar to the IDM, but 
instead of imposing an exact $Z_2$ symmetry, the model allows softly broken $Z_2$ symmetric terms,
which still avoids flavor-changing neutral currents at tree level~\cite{Glashow:1976nt}.
The Higgs potential in the THDM is given by
\begin{align}
V(\Phi_1,\Phi_2) &= m_1^2|\Phi_1|^2+m_2^2|\Phi_2|^2-(m_3^2\Phi_1^\dagger \Phi_2^{} +\text{h.c.})\notag\\
&\quad +\frac{1}{2}\lambda_1|\Phi_1|^4+\frac{1}{2}\lambda_2|\Phi_2|^4+\lambda_3|\Phi_1|^2|\Phi_2|^2+\lambda_4|\Phi_1^\dagger\Phi_2^{}|^2
+\frac{1}{2}\{\lambda_5(\Phi_1^\dagger\Phi_2^{})^2+\text{h.c.}\},
\label{pot_thdm}
\end{align}
where $m_3^2$ and $\lam_5$ can be real by assuming the $CP$ conservation.
The two doublet fields $\Phi_{1,2}$ are parameterized as 
\begin{align}
\Phi_i=\left(\begin{array}{c}
w_i^+\\
\frac{1}{\sqrt{2}}(v_i+h_i+iz_i)
\end{array}\right)~~\text{with}~~i=1,2,
\end{align} 
where $v_1$ and $v_2$ are the VEVs of the Higgs doublet fields with $v=(v_1^2+v_2^2)^{1/2}$.
The mass eigenstates of the Higgs fields are defined as
\begin{align}
\left(\begin{array}{c}
h_1\\
h_2
\end{array}\right)=R(\alpha)
\left(\begin{array}{c}
H\\
h
\end{array}\right),
\quad
\left(\begin{array}{c}
z_1\\
z_2
\end{array}\right)
=R(\beta)\left(\begin{array}{c}
G^0\\
A
\end{array}\right),
\quad
\left(\begin{array}{c}
w_1^\pm\\
w_2^\pm
\end{array}\right)&=R(\beta)
\left(\begin{array}{c}
G^\pm\\
H^\pm
\end{array}\right),
\label{mixing}
\end{align}
where $R(\theta)$ is a rotation matrix with a mixing parameter $\theta$ and
$\tan\beta = v_2/v_1$. 
After imposing two stationary conditions for $h_{1,2}$, 
masses of the physical Higgs bosons and the mixing angle $\alpha$ are expressed by 
\begin{align}
&m_{H^\pm}^2 = M^2-\frac{1}{2}v^2(\lambda_4+\lambda_5),\quad \notag
 m_A^2=M^2-v^2\lambda_5,  \\ 
&m_H^2=M_{11}^2 c^2_{\beta-\alpha} + M_{22}^2 s^2_{\beta-\alpha} -M_{12}^2 s_{2(\beta-\alpha)}, \quad 
m_h^2=M_{11}^2 s^2_{\beta-\alpha}  + M_{22}^2 c^2_{\beta-\alpha}  +M_{12}^2 s_{2(\beta-\alpha)}, \notag \\
&\tan 2(\beta-\alpha)= -\frac{2M_{12}^2}{M_{11}^2-M_{22}^2}, \label{333}
\end{align} 
where $s_\theta$ and $c_\theta$ represent $\sin\theta$ and $\cos\theta$, 
$M^2 \equiv m_3^2/s_\beta c_\beta$ describes the soft breaking scale of the $Z_2$ symmetry, and $M_{ij}^2$ are the mass matrix elements for the $CP$-even scalar states in the basis of $(h_1,h_2)R(\beta)$:
\begin{align}
&M_{11}^2=v^2(\lambda_1c^4_\beta+\lambda_2s^4_\beta+2\lambda_{345}^+s^2_{\beta}c^2_\beta),\quad
M_{22}^2=M^2+\frac{1}{4}v^2s^2_{2\beta}(\lambda_1+\lambda_2-2\lambda_{345}^+), \notag\\
&M_{12}^2=\frac{1}{2}v^2 s_{2\beta}( -\lambda_1c^2_\beta+\lambda_2s^2_\beta+\lambda_{345}^+ c_{2\beta}  ).
\end{align}
We choose the following free parameters in the THDM:
\begin{align}
 m_H^{},~ m_A^{},~ m_{H^+},~ M^2,~ \tan\beta,~ c_{\beta-\alpha}.
\end{align} 

Unlike the inert models, additional Higgs bosons in the THDM couple with SM fermions through the Yukawa interaction.
We can define four types of interactions under the softly broken $Z_2$ symmetry:
Type-I, II, X and Y, depending on the $Z_2$ charge assignment for the right-handed fermions~\cite{Barger:1989fj,Grossman:1994jb,Akeroyd:1994ga,Akeroyd:1996he,Akeroyd:1998ui,Aoki:2009ha}.
In all four types of THDMs, the up-type Yukawa interaction is modified in the same manner.
In the Type-II and Type-Y THDMs, on the other hand, the down-type Yukawa can be enhanced by $\tan\beta$.

The tree-level couplings of the SM-like Higgs boson to SM fermions and weak bosons are modified via the mixing with the other neutral Higgs boson. 
The scaling factors $\kappa_f$ and $\kappa_V$ are
\begin{align}
 \kappa_f=s_{\beta-\alpha}+\zeta_f c_{\beta-\alpha}, \quad \kappa_V=s_{\beta-\alpha},
 \label{kappa_thdm}
\end{align}
where $\zeta_f=\cot\beta$ for the up-type Yukawa coupling.
The $h$--$H^+$--$H^-$ coupling is given by
\begin{align}
 g_{hH^+H^-}
  =-\frac{2}{v}\Big\{\Big(m_{H^+}^2-M^2+\frac{1}{2}m_h^2\Big)\sba
      -(M^2-m_h^2)\cot2\beta\,\cba\Big\}
  \equiv -v\lambda'_3. 
\label{ghHH_thdm}
\end{align}
For later comparison, $\lambda'_3$ is defined in the same manner as in Eq.~\eqref{ghHH_idm} in the IDM.

\begin{figure}
 \includegraphics[height=0.5\textwidth]{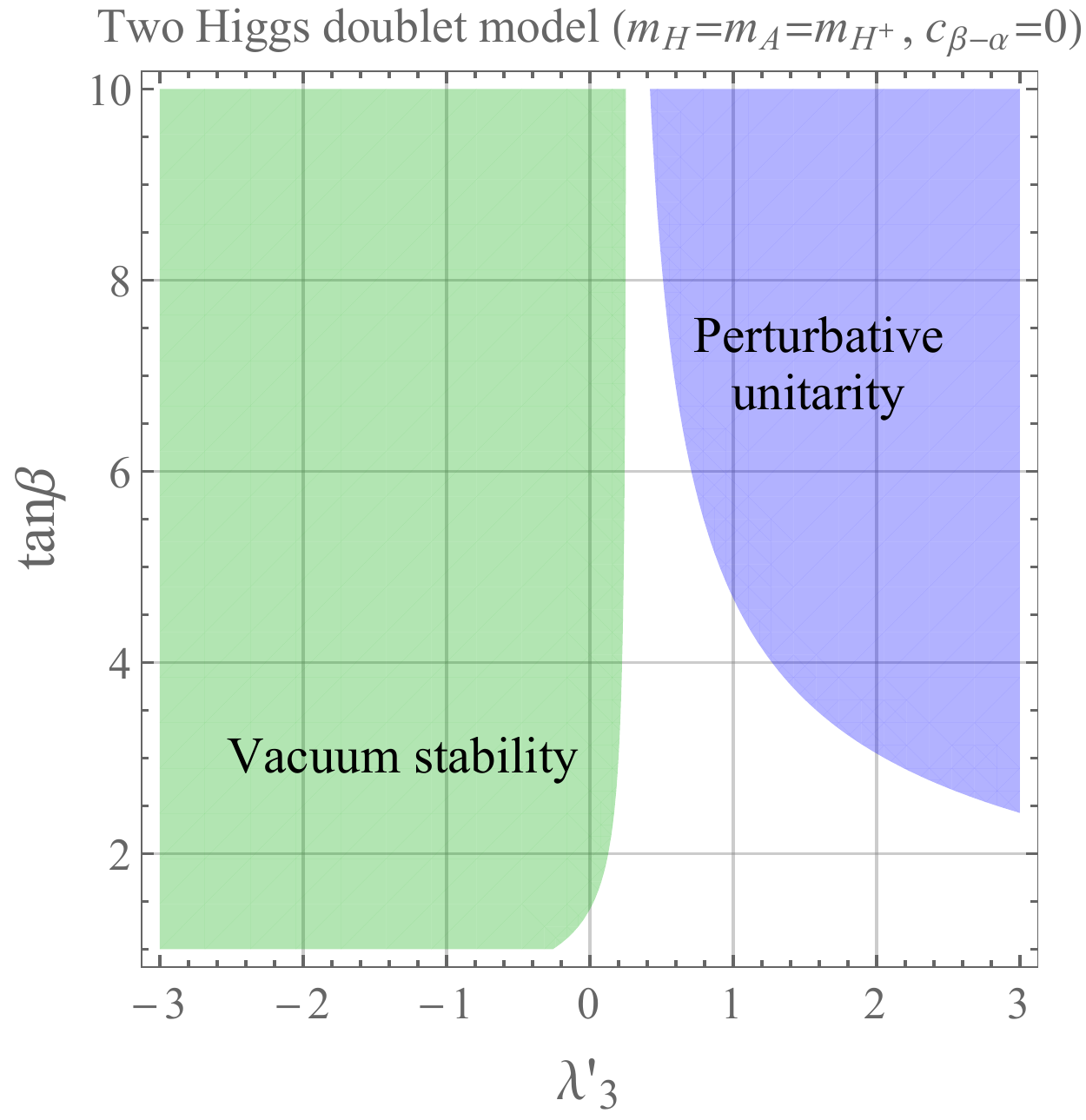}
 \caption{Theoretical constraints on the $\lam'_3$--$\tan\beta$ plane in the THDM.}
\label{fig:thconst_thdm}
\end{figure}

Regions of the parameter space can be constrained by imposing the bounds from 
vacuum stability~\cite{Deshpande:1977rw,Sher:1988mj,Kanemura:1999xf},
perturbative unitarity~\cite{Kanemura:1993hm,Akeroyd:2000wc,Ginzburg:2005dt,Kanemura:2015ska},
and the $S$ and $T$ parameters~\cite{Toussaint:1978zm,Bertolini:1985ia,Peskin:2001rw,Grimus:2008nb,Kanemura:2011sj}. 
Figure~\ref{fig:thconst_thdm} shows allowed regions of $\lambda'_3$ as a function of $\tan\beta$
in the THDM after imposing the above theoretical constraints, where we take $c_{\beta-\alpha}=0$
and $m_H=m_A=m_{H^+}$ for an illustration.

Because of the Yukawa interactions of additional Higgs bosons, 
the parameter space is additionally constrained by the LHC experiments, which
are summarized in Ref.~\cite{Arbey:2017gmh}.
The mass below 160~GeV is excluded from $t\to H^\pm b$, $H^\pm\to\tau\nu$ searches
for $\tan\beta<4,8,10$ in the Type-Y, Type-I, and Type-X models, respectively. 
In the Type-II model, on the other hand, the same mass region is excluded irrespective of $\tan\beta$.
Moreover, flavor experiments such as $B$ meson decays also give an important constraint particularly on the mass of the charged Higgs bosons and $\tan\beta$~\cite{Misiak:2017bgg}.
In the Type-II (or Type-Y) model, $m_{H^+}>580$~GeV independently of $\tan\beta$.
In the Type-I (or Type-X) model, $m_{H^+}>445$~GeV for $\tan\beta=1$, while there is no constraint for $\tan\beta>2$.
As we see later, for $e^+e^-\to h\gamma$, only light charged scalars are relevant, so we only consider the Type-I (or equivalently Type-X) THDM below.

\section{Numerical results}\label{sec:results}

In this section, we describe the formalism of our calculations,
and show the numerical results at $\sqrt{s}=250$~GeV as well as at $\sqrt{s}=500$~GeV.

\subsection{Loop calculations by using the {\sc H-COUP} program}\label{sec:hcoup}

We consider the process
\begin{align}
 e^-(k,\sig/2) +e^+(\bar k,\bar\sig/2)\to \gamma(p,\lam) + h(p_h),
\end{align}
where the four-momenta ($k,\bar k,p,p_h$) and helicities ($\sig,\bar\sig,\lam$)
are defined in the center-of-mass (CM) frame of the collision.
The amplitudes are given by 
\begin{align}
 {\cal M}_{\sig\bar\sig}^{\lam}
 =\bar v(\bar k,\bar\sig)\,\Gamma^\nu\, u(k,\sig)\,\epsilon_\nu^*(p,\lam),
\end{align}
where the vertex function $\Gamma^\nu$ can be decomposed in terms of the form factors $C_i^\sig$ as~\cite{Djouadi:1996ws} 
\begin{align}
 \Gamma^\nu=\gamma_\mu \big\{C_1^\sigma\big(g^{\mu\nu}p\cdot k-p^\mu k^\nu\big)
 +C_2^\sigma\big(g^{\mu\nu}p\cdot \bar k-p^\mu \bar k^\nu\big)
 +C_3^\sigma g^{\mu\nu}   \big\}.
\end{align}
$C_i^\sig$ sum all the one-loop contributions as
\begin{align}
 C_i^\sig = C_i^{hee,\sig}+C_i^{hV\gamma,\sig}+C_i^{{\rm box},\sig},
\end{align}
where $C_i^{hee,\sig}$ denotes the $hee$ vertex contributions,
$C_i^{hV\gamma,\sig}$ ($V=\gamma,Z$) are the triangle vertex contributions 
 (diagrams (a,b,d) in Fig.~\ref{fig:diagram}), 
 and $C_i^{{\rm box},\sig}$ is the box contributions (diagram (c)).
Note that, due to the gauge invariance, $C_3^\sig$ is zero after summing all the contributions. 
The differential cross section is~\cite{Sang:2017vph} 
\begin{align}
 \frac{d\sig}{d\cos\theta}
 =\frac{s-m_h^2}{64\pi s}\big\{t^2\big(|C_1^+|^2+|C_1^-|^2\big)
 +u^2\big(|C_2^+|^2+|C_2^-|^2\big)\big\},
\end{align}
where $\theta$ is the scattering angle between the electron and the photon in the CM frame,
and $t,u=-(s-m_h^2)(1\mp\cos\theta)/2$.
In the following, we briefly describe how we evaluate the form factors $C_i^\sig$ for each loop diagram by using the {\sc H-COUP} program~\cite{Kanemura:2017gbi}.

{\sc H-COUP} is a numerical code to evaluate the renormalized gauge-invariant vertex functions  for the SM-like Higgs boson at one-loop level
in various extended Higgs models~\cite{Kanemura:2017gbi}.
So far, the Higgs singlet model (HSM), four types of the THDMs and the IDM are included, 
based on Refs.~\cite{Kanemura:2004mg,Kanemura:2014dja,Kanemura:2015mxa,Kanemura:2015fra,Kanemura:2016sos,Kanemura:2016lkz,Kanemura:2017wtm}. 
Recently the program was extended to include box contributions in order to evaluate 
the three-body decay $h\to Zf\bar f$ at the one-loop level~\cite{Kanemura:2018yai}. 
Since the $h\to Zf\bar f$ decay is related to the $e^+e^-\to h\gamma$ process by crossing symmetry and by replacing the $Z$ boson by the photon, 
we utilize the program with slight extensions to evaluate the form factors $C_i^\sig$ both in the SM and in extended Higgs models. 
For the ITM we newly implemented the relevant pieces, based on the IDM.
We validated our numerical calculations by comparing with the earlier works in the SM~\cite{Djouadi:1996ws,Sang:2017vph} as well as in the IDM~\cite{Arhrib:2014pva},
while the results in the ITM and the THDMs are reported for the first time in this article.

In the $m_e=0$ limit,
the one-loop $hee$ vertex in extended Higgs models deviates from the SM prediction only by the mixing effects as the $\kappa_V$ scaling factor due to weak-boson loops.  
However,
the $hee$ vertex diagrams are nonzero only for $C_3^\sig$; i.e., do not contribute to the physical observable.
In order to check the gauge invariance, we evaluate $C_3^{hee,\sig}$ by using {\sc H-COUP}.
Similarly, the box contributions in extended Higgs models are modified only by the $\kappa_V$ factor.
See Ref.~\cite{Djouadi:1996ws} for the explicit formulas in the SM.

The most important contribution in extended Higgs models comes from the triangle loop diagrams, especially
charged scalar loops.
The renormalized $hV\gamma$ vertices are decomposed by the following two form factors in {\sc H-COUP},
\begin{align}
{\Gamma}_{hV\gamma}^{\mu\nu}(p^2,q^2,p_h^2)
 =g^{\mu\nu}{\Gamma}_{hV\gamma}^1 -p^\mu q^\nu {\Gamma}_{hV\gamma}^2 
\label{hvv}
\end{align}
with $q=k+\bar k$ being the momentum of the $s$-channel off-shell photon or $Z$ boson.
As shown in Fig.~\ref{fig:diagram},
${\Gamma}_{hV\gamma}^{\mu\nu}$ includes top and $W$ loops with the scaling factors $\kappa_f$ and $\kappa_V$ as well as charged scalar loops in extended Higgs models.
$C_{1,2,3}^{hV\gamma,\sig}$ can be evaluated in terms of $\Gamma_{hV\gamma}^{1,2}$ as
\begin{align}
 C_1^{hV\gamma,\sig}=C_2^{hV\gamma,\sig}
  =-\frac{g^\sig_{Vee}}{s-m_V^2}{\Gamma}_{hV\gamma}^2,\quad
 C_3^{hV\gamma,\sig}
   =-\frac{g^\sig_{Vee}}{s-m_V^2}\big({\Gamma}_{hV\gamma}^1 -p\cdot q\, {\Gamma}_{hV\gamma}^2\big),
\label{C123}
\end{align}  
where $g^\sig_{Vee}$ are the SM $\gamma ee$ and $Zee$ couplings.
One can find the explicit expressions for $\Gamma_{h\gamma\gamma}^{1,2}$ and $\Gamma_{hZ\gamma}^{1,2}$ in  the Appendix.
For the validity of our program,
we numerically check the gauge invariance; i.e., ${\Gamma}_{hV\gamma}^1=p\cdot q\, {\Gamma}_{hV\gamma}^2$ for top and charged scalar loops independently, 
and $C_3^\sig=0$ for weak-boson loops after summing all the contributions.
When $q^2=0$ or $m_Z^2$, namely, the case for the $\haa$ or $h\to Z\gamma$ decay,
we can also check ${\Gamma}_{hV\gamma}^1=p\cdot q\, {\Gamma}_{hV\gamma}^2$ even for 
$W$-boson loops.

\subsection{Results at $\sqrt{s}=250$~GeV}

We show numerical results for the cross sections at $\sqrt{s}=250$~GeV in the IDM, the ITM, and the THDM, in order. 
We also discuss correlations between the $\eeha$ production and the $\haa$ decay
as well as the $\hza$ decay.
In order to see deviations from the predictions in the SM, we evaluate the ratios
of the total cross sections and the partial decay rates
\begin{align}
 \Delta R(e^+e^-\to h\gamma)&=\frac{\sigma_{\rm NP}(e^+e^-\to h\gamma)}{\sigma_{\rm SM}(e^+e^-\to h\gamma)}-1 &&\hspace*{-3cm}(\equiv\Delta R_{h\gamma}),
 \label{delRp} \\
\Delta R(h\to V\gamma)&=\frac{\Gamma_{\rm NP}(h\to V\gamma)}{\Gamma_{\rm SM}(h\to V\gamma)}-1 &&\hspace*{-3cm}(\equiv\Delta R_{V\gamma}), 
\label{delRd} 
\end{align} 
where $V=\gamma$ or $Z$, and 
$\sigma_{\rm NP}$ and $\Gamma_{\rm NP}$ are evaluated in each extended Higgs model.

\paragraph{IDM:}

In the IDM, the relevant parameters for the $\eeha$ process are the charged Higgs-boson mass $\mc$ and the $h$--$H^+$--$H^-$ coupling $g_{hH^+H^-}$ in Eq.~\eqref{ghHH_idm}.
We choose $\mc$ and $\lam_3$ as free parameters for our illustration,
while we fix the other parameters, $m_H$, $m_A$ and $\lam_2$, so as to avoid the constraints discussed in Sec.~\ref{sec:idm}.%
\footnote{For instance, with $m_H=62.5$~GeV, $\mc=m_A=100$~GeV and $\lambda_2=8.25$,
the range $\lam_3\gtrsim-1.2$ is allowed. We note that such a parameter region also satisfies dark matter constraints~\cite{Belyaev:2016lok,Ilnicka:2018def}. 
}
We note that our results agree with the previous study in Ref.~\cite{Arhrib:2014pva}.

\begin{figure}
 \includegraphics[height=0.4\textheight]{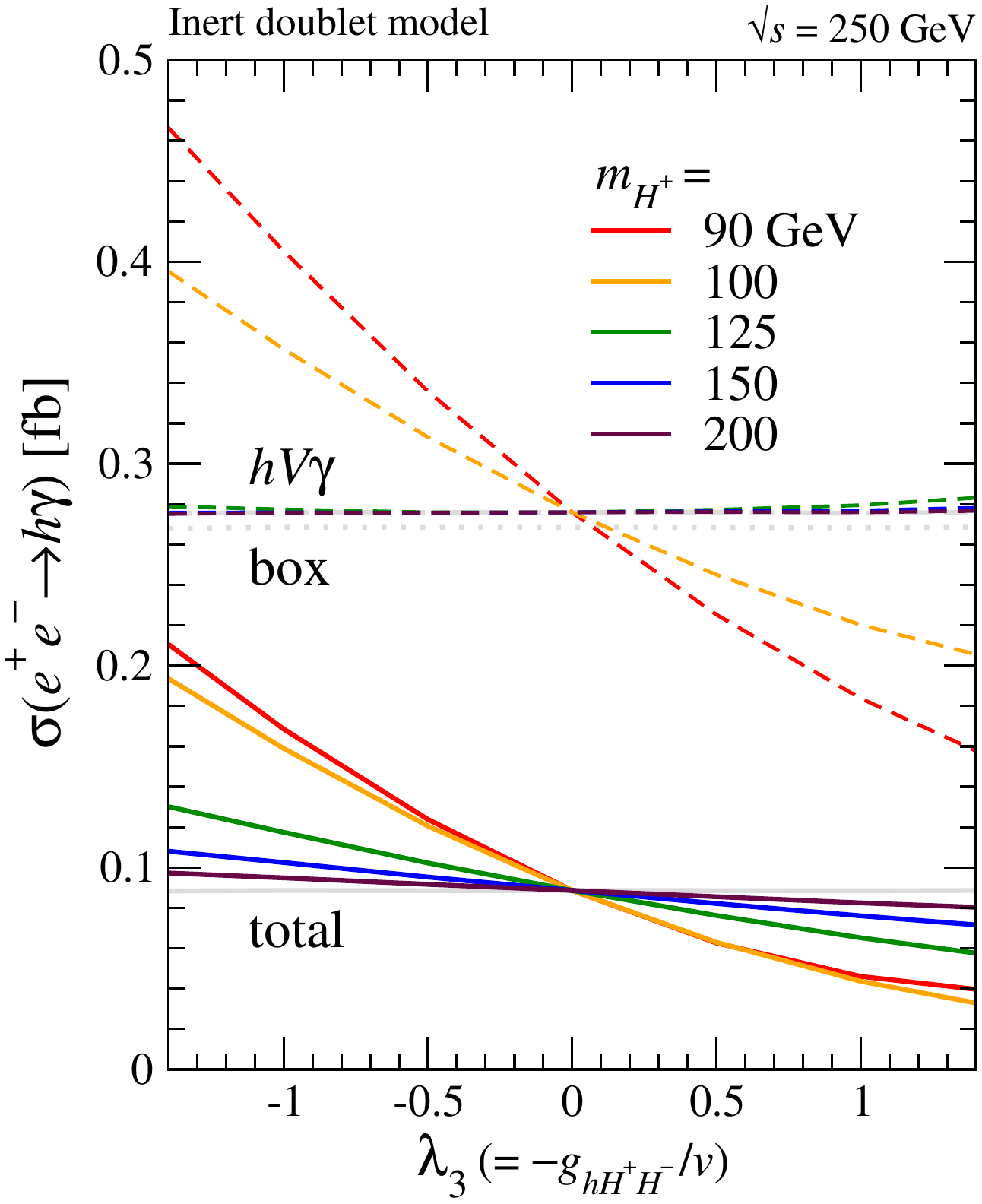}\qquad
 \includegraphics[height=0.4\textheight]{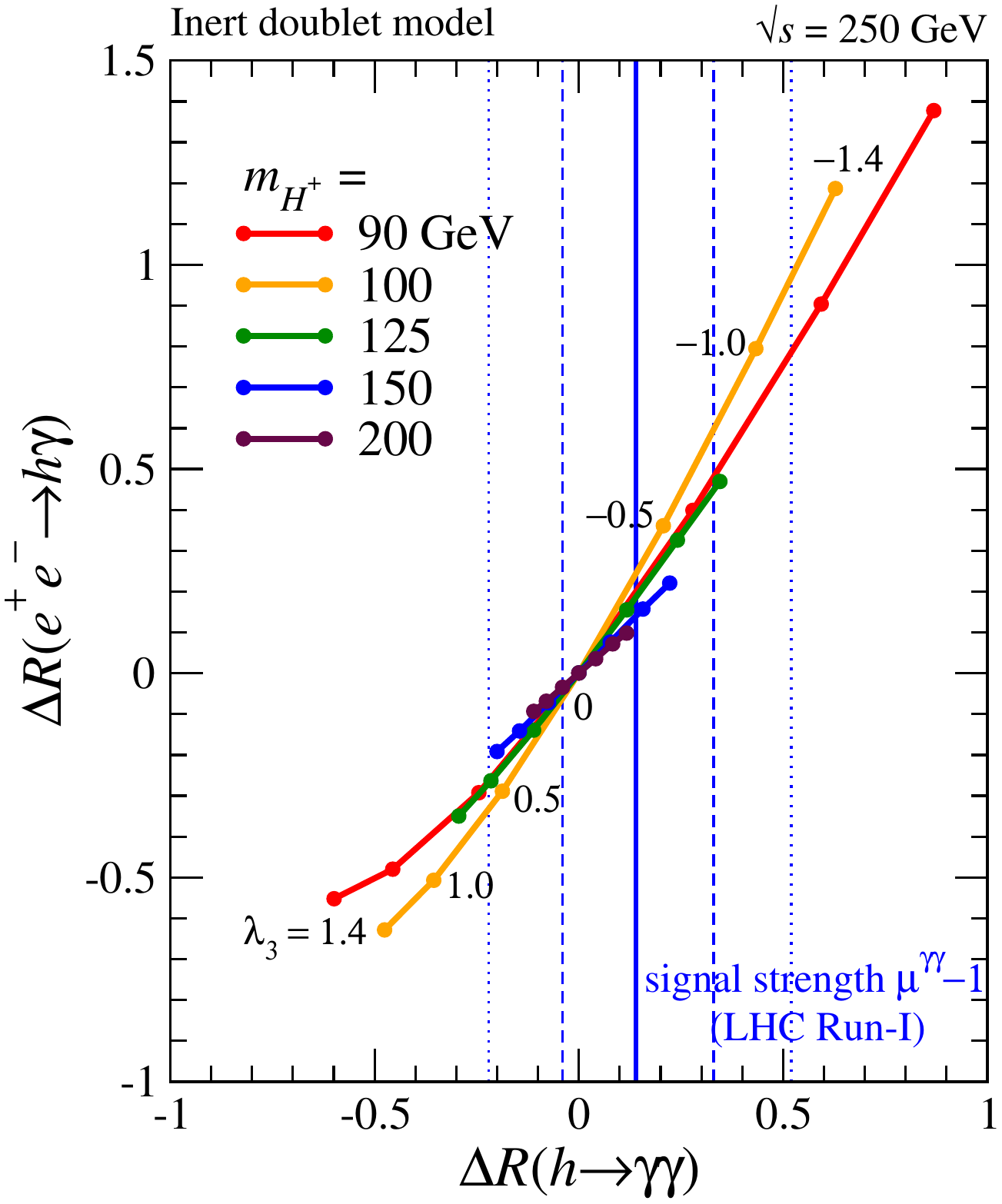}
 \caption{
 [Left] Total cross sections for $\eeha$ in the IDM as a function of $\lam_3$  
 at $\sqrt{s}=250$~GeV with different charged Higgs-boson masses.
 The contributions only from the $hV\gamma$ (dashed lines) or box (dotted) diagrams are also shown as a reference.
 Gray lines denote the SM prediction.
 [Right] Correlations between $\Delta R(\haa)$ and $\Delta R(\eeha)$, 
 defined in Eqs.~\eqref{delRd} and \eqref{delRp}.
 Vertical blue lines indicate the $\haa$ signal strength from the LHC Run-I data with the 1$\sig$ and 2$\sig$ uncertainties.
 }
\label{fig:rs250_idm}
\end{figure}

In Fig.~\ref{fig:rs250_idm}(left),
we show total cross sections for $\eeha$ in the IDM as a function of $\lam_3$.
In addition to the total cross sections (solid lines), we separately show the triangle $hV\gamma$ (dashed) and box (dotted) contributions as a reference.
Horizontal gray lines denote the SM prediction.
In the IDM, as shown in Eq.~\eqref{kappa_idm},
the SM-like Higgs-boson couplings to SM fermions and weak bosons are identical to the SM ones.
Therefore, the box contributions are same as in the SM.
On the other hand, the top and $W$ loops in the $hV\gamma$ contributions are also same as in the SM, but
the charged scalar loops give additional contributions and modify the total rate if $g_{hH^+H^-}$ is nonzero and the charged Higgs boson is light enough. 
At $\sqrt{s}=250$~GeV, if $\mc<125$~GeV; i.e., above the $H^+H^-$ pair threshold, 
the amplitudes of the charged scalar loops are complex, while those are real for $\mc\geq125$~GeV;
see Fig.~\ref{fig:loopfunc} in the Appendix.
This explains the quantitative difference for the deviations between the cases below and above 
$\mc=\sqrt{s}/2=125$~GeV.

When $\lam_3$ is negative (positive), the production is enhanced (reduced)
from the SM prediction.
Because of $\lam_3\gtrsim-1.4$ for $\lam_2\sim8$ from the theoretical constraints (see Fig.~\ref{fig:thconst}(left))
and because of the exclusion bound of $\mc$
from the null result of the direct search at the LEP, 
as discussed in Sec.~\ref{sec:idm},
a possible largest enhancement in the IDM is about 2.5 times the SM prediction.

The same triangle $H^\pm$ loops appear in the $\haa$ process, which can modify the observed Higgs decay rate.
One can find the explicit partial decay rate in the IDM in Refs.~\cite{Cao:2007rm,Arhrib:2012ia,Swiezewska:2012eh,Krawczyk:2013jta},
which is implemented in {\sc H-COUP}~\cite{Kanemura:2016sos}.
In Fig.~\ref{fig:rs250_idm}(right), we show correlations
between $\draa$ and $\drha$, 
defined in Eqs.~\eqref{delRd} and \eqref{delRp}, respectively.
There are clear positive correlations. 
As the $\eeha$ process is more intricate than $\haa$, 
namely, the off-shellness of one of the photons and the additional $hZ\gamma$ and box contributions,
the slope of the correlations differs by the charged Higgs-boson mass; 
$|\drha|>|\draa|$ for $\mc<150$~GeV, 
while $\drha\sim\draa$ for $\mc\gtrsim150$~GeV.

In Fig.~\ref{fig:rs250_idm}(right), we also present the $\haa$ signal strength
from the LHC Run-I data with the 1$\sig$ and 2$\sig$ uncertainties,
$\mu^{\gamma\gamma}=1.14^{+0.19}_{-0.18}$~\cite{Khachatryan:2016vau}, by vertical blue lines.
This provides the stronger constraints for light charged Higgs bosons.
For instance, for the case of $\mc=100$~GeV, $-1.2\lesssim\lam_3\lesssim0.6$ is allowed at 95\% confidence level (CL).
If we observe $\draa\sim0$ with smaller uncertainties in future experiments,
the parameter region of large $|\lam_3|$ and small $\mc$ will be further constrained.

\paragraph{ITM:}

\begin{figure}
 \center 
 \includegraphics[height=0.4\textheight]{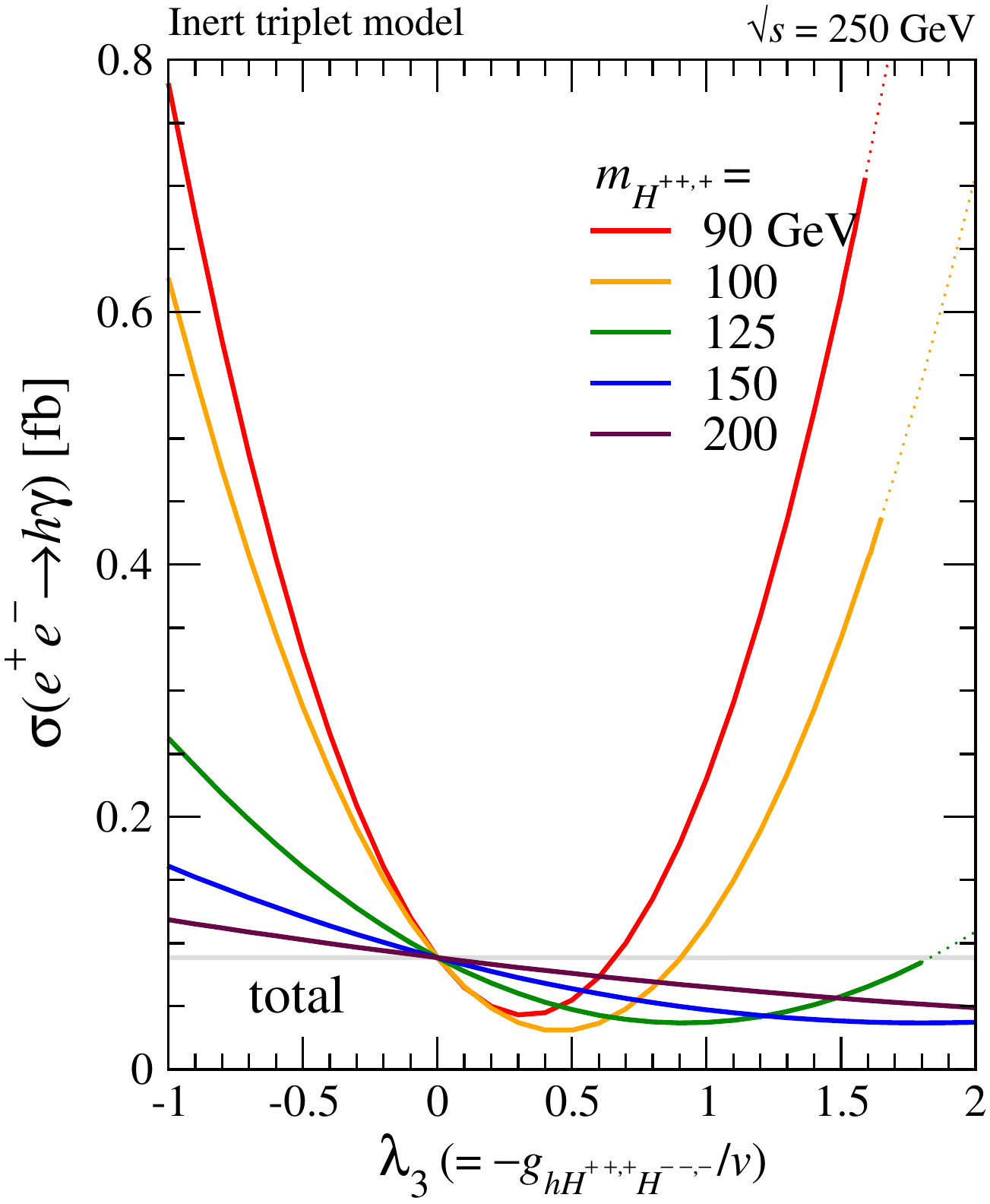}\qquad
 \includegraphics[height=0.4\textheight]{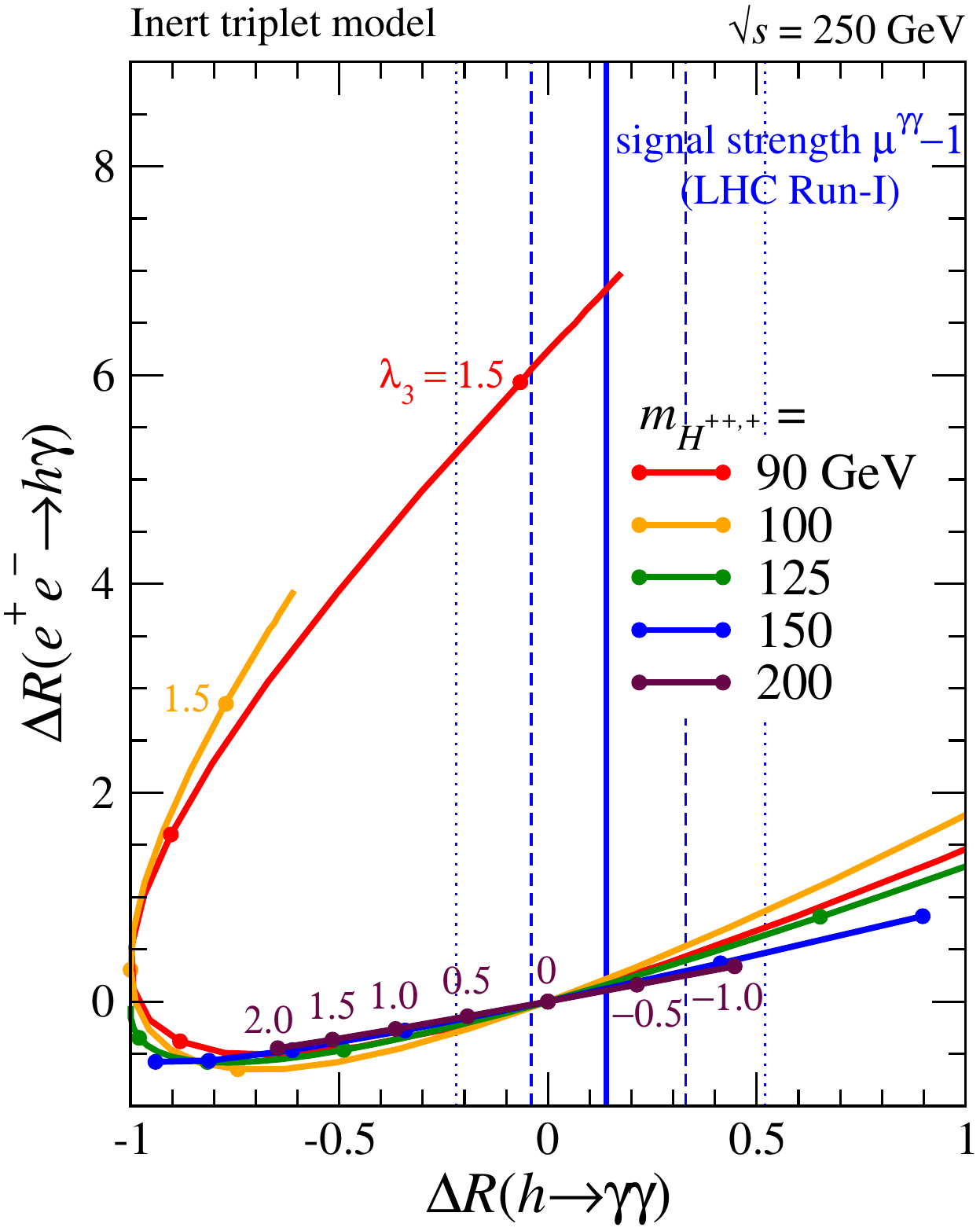}
 \caption{
 [Left] Total cross sections for $\eeha$ in the ITM as a function of $\lam_3$  
 at $\sqrt{s}=250$~GeV with different charged Higgs boson masses.
 The right regions shown by dotted lines for $m_{H^{++}}=m_{H^{+}}=90, 100, 125$~GeV indicate parameter regions excluded by the requirement of the existence of the inert vacuum.
 A gray line denotes the SM prediction.
 [Right] Correlations between $\Delta R(\haa)$ and $\Delta R(\eeha)$. 
 Vertical blue lines indicate the $\haa$ signal strength from the LHC Run-I data with the 1$\sig$ and 2$\sig$ uncertainties. 
 }
\label{fig:rs250_itm}
\end{figure}

Turn to the ITM.
For simplicity, we assume $\mcc=m_H$ (or $\lam_5=0$).
Under this assumption,   
$g_{hH^{++}H^{--}}=g_{hH^+H^-}=-v\lambda_3$ and $\mc=\mcc$,
and hence the situation is very similar to the IDM, except for additional loops of the doubly charged scalars. 

Similarly to Fig.~\ref{fig:rs250_idm}(left) in the IDM, we show total cross sections in the ITM as a function of $\lam_3$
in Fig.~\ref{fig:rs250_itm}(left).
The production rate is highly enhanced 
due to the considerable contributions from the doubly charged scalars.  
This is simply because photons couple to particles with the electromagnetic charge.
The vertex functions $\Gamma_{h\gamma\gamma}^{1,2}$ in Eqs.~\eqref{Ghaa1} and \eqref{Ghaa2} are proportional to $\sum_S Q_S^2$, 
and hence, for the $\lam_5=0$ case, 
 the contribution from the charged scalar loops of the $h\gamma\gamma$ vertex in the ITM is five times larger than that in the IDM at the amplitude level. 
 
A remarkable difference from the IDM is that, even for positive $\lam_3$, the production rate can be significantly enhanced if the charged Higgs bosons are light. 
This is because the charged scalar contributions overwhelm all the other contributions. 
A caveat is that there is an upper bound on $\lam_3$ from the requirement of the existence of the inert vacuum,
which depends on the masses of the inert scalars, as shown in Fig.~\ref{fig:thconst}(right).
The excluded regions are indicated by dotted lines. 
We also note that, even in the IDM, we naively expect that the cross section could be enhanced
if we take much larger positive $\lam_3$.
However, again, such parameter region is not allowed by the theoretical constraints. 

In Fig.~\ref{fig:rs250_itm}(right), we show correlations between $\draa$ and $\drha$. 
The $h\to\gamma\gamma$ decay with doubly charged scalars has been studied
in Refs.~\cite{Alves:2011kc,Arhrib:2011vc,Kanemura:2012rs,Akeroyd:2012ms,Chiang:2012qz,Chun:2012jw,Aoki:2012jj,Dev:2013ff,Dey:2018uvu},
and was newly implemented in {\sc H-COUP}.
The lower part of the plot ($\drha<2$) is qualitatively similar to Fig.~\ref{fig:rs250_idm}(right) in the IDM.
Because of the huge enhancement (reduction) of the production and the decay by the doubly charged scalars,   
the LHC $\haa$ signal strength provides severer constraints on the masses and the couplings than in the IDM.
For instance, for the case of $m_{H^{++}}=m_{H^{+}}=100\ (200)$~GeV,
$-0.2\ (-1.1)\lesssim\lam_3\lesssim0.1\ (0.6)$
is allowed at 95\% CL.

Again, a remarkable difference from the IDM is that 
we can find a particular parameter region where $\sigma(\eeha)$ can be significantly enhanced by a factor of 
$6-8$, but $\Gamma(\haa)$ still remains as in the SM; e.g.
the case for $m_{H^{++}}=m_{H^{+}}=90$~GeV with 
 $1.4\lesssim\lam_3\lesssim1.6$ is allowed at 95\% CL under the theoretical and current experimental constraints. 
Such parameter regions are quite attractive for future $e^+e^-$ colliders,
in which we can expect a synergy between the observation of the rare $h\gamma$ production 
and the direct discovery of charged Higgs bosons.

\paragraph{THDM:}

\begin{figure}
 \center 
 \includegraphics[height=0.4\textheight]{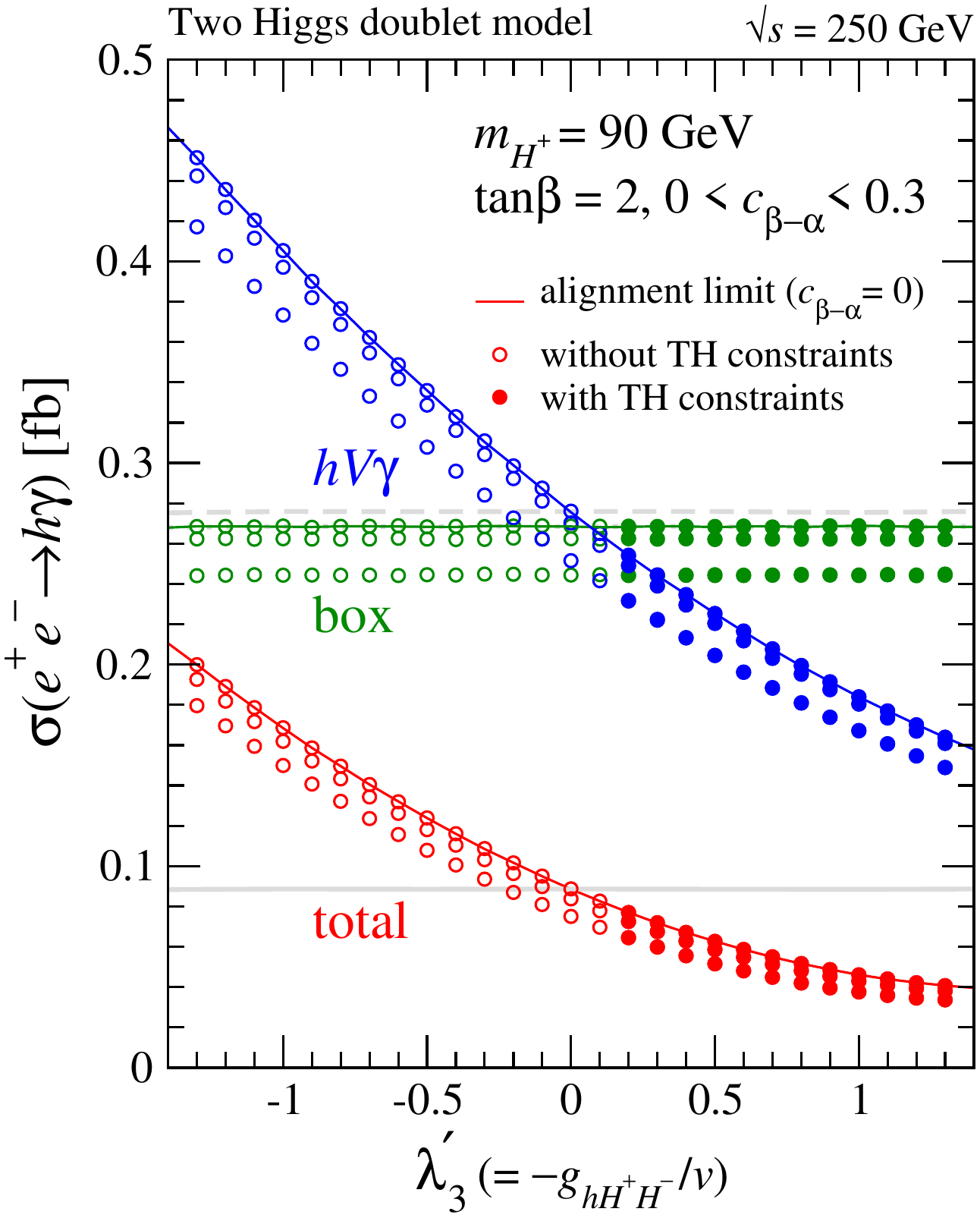}\qquad
 \includegraphics[height=0.4\textheight]{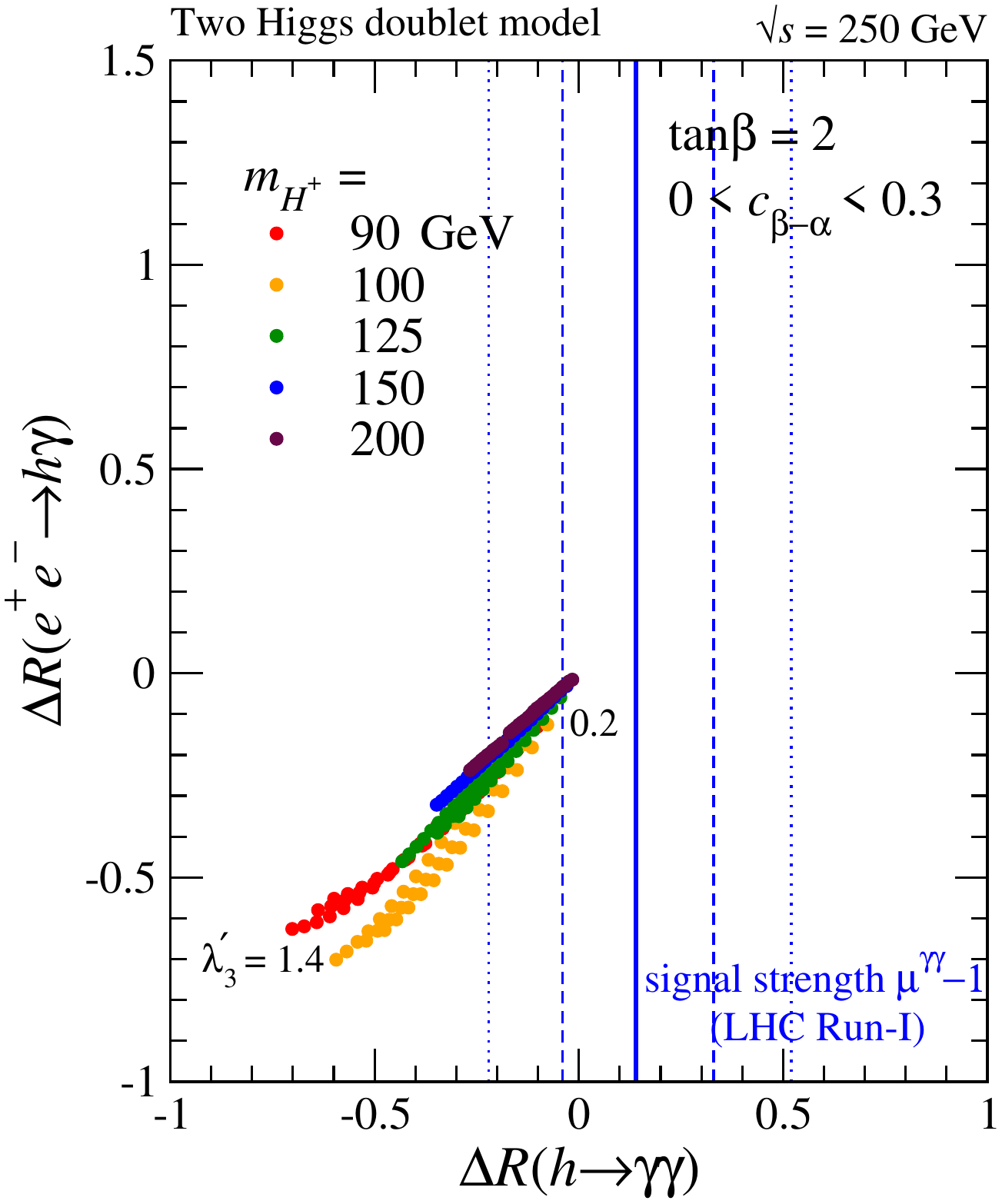}
 \caption{
 [Left] Total cross sections for $\eeha$ in the THDM as a function of $\lam_3$  
 at $\sqrt{s}=250$~GeV for $\mc=90$~GeV,
 where we fix $\tan\beta=2$ and vary $\cba$ from 0 to 0.3.
 Gray lines denote the SM prediction.
 [Right] Correlations between $\Delta R(\haa)$ and $\Delta R(\eeha)$. 
 Vertical blue lines indicate the $\haa$ signal strength from the LHC Run-I data with the 1$\sig$ and 2$\sig$ uncertainties. 
 }
\label{fig:rs250_thdm}
\end{figure}

In the THDM with softly broken $Z_2$ symmetry, the couplings of the SM-like Higgs boson with SM fermions and weak bosons are modified through the mixing with other neutral scalars, as shown in Eq.~\eqref{kappa_thdm}.
For $\cba=0$, the so-called alignment limit, there is no mixing effect; i.e., $\kappa_f=\kappa_V=1$. 
Therefore, if we take the same value of $g_{hH^+H^-}$ in Eq.~\eqref{ghHH_thdm}; i.e., $\lam_3'=\lam_3$,
the prediction in the THDM is identical to that in the IDM, which is explicitly shown
for $\mc=90$~GeV by solid lines in Fig.~\ref{fig:rs250_thdm}(left).

Once $\cba$ deviates from zero, the scaling factors $\kappa_{f}$ and $\kappa_V$ deviate from one. 
In order to see this mixing effect, we take $\tan\beta=2$ and vary $\cba$ from 0 to 0.3, which is allowed 
in the current constraints from the Higgs coupling measurements~\cite{Aad:2015pla,CMS:2016qbe}.
We note that a negative $\cba$ is also allowed, and the results are very similar to the positive case.
Unlike the inert models, not only the $hV\gamma$ contributions but also
the box contributions deviate from the SM due to the scaling factors of the SM-like Higgs couplings.
Therefore, even for $\lam'_3=0$, the cross section in the THDM differs from that in the SM if $\cba\ne0$.
Except such small mixing effects, the qualitative behaviors are very similar to the case in the IDM,
namely negative (positive) $\lam'_3$ can enhance (reduce) the production rate.

As shown in filled circles in Fig.~\ref{fig:rs250_thdm}(left),
however, the theoretical constraints in the THDM are much severer than those in the IDM;
the vacuum stability requires $\lam'_3>0.1$ for $\tan\beta>2$ (see Fig.~\ref{fig:thconst_thdm}).
As mentioned in Sec.~\ref{sec:thdm}, although such light charged Higgs bosons with $\tan\beta=2$ are allowed 
in the Type-I and Type-X models for the $B$-physics experiments~\cite{Misiak:2017bgg},
the current LHC data require higher $\tan\beta$ values to avoid the lower limit on the mass,
$\mc>160$~GeV~\cite{Arbey:2017gmh}.
On the other hand, for larger $\tan\beta$, the allowed region with the theoretical constraints in Fig.~\ref{fig:rs250_thdm}(left) becomes smaller due to the unitarity constraint.%
\footnote{For instance, for $m_{H^+}=90$~GeV with $\tan\beta=8$, which is still allowed by experiments in the Type-I THDM, $\lambda'_3$ is constrained between 0.3 and 0.5 by the vacuum stability and the unitarity; see also Fig.~\ref{fig:thconst_thdm}.}
Therefore, in the THDMs, there is a tension between the allowed parameter region and the preferred region to observe large deviations in the $\eeha$ signal.  

\begin{figure}
 \includegraphics[width=0.45\textwidth]{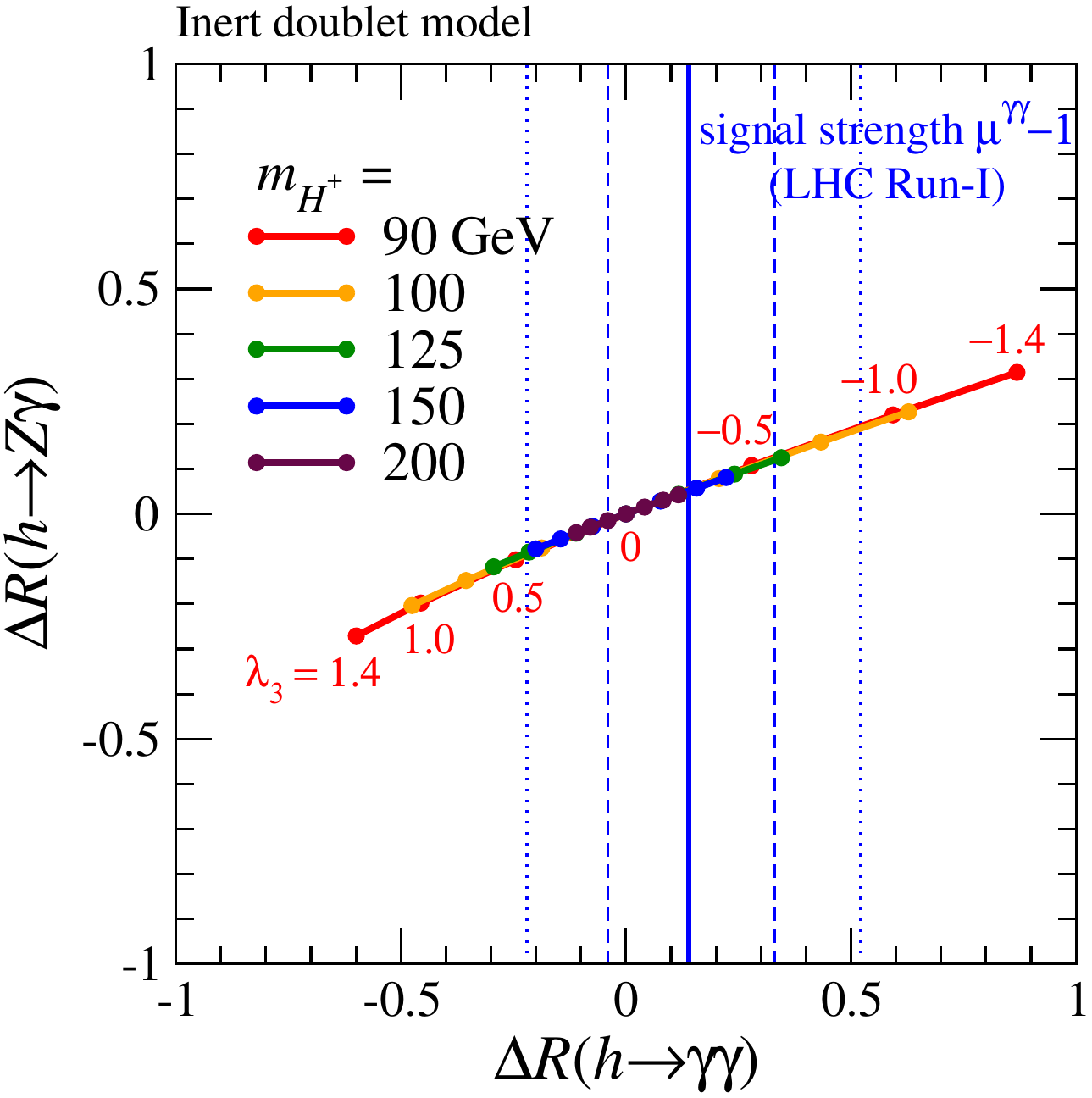}\qquad
 \includegraphics[width=0.45\textwidth]{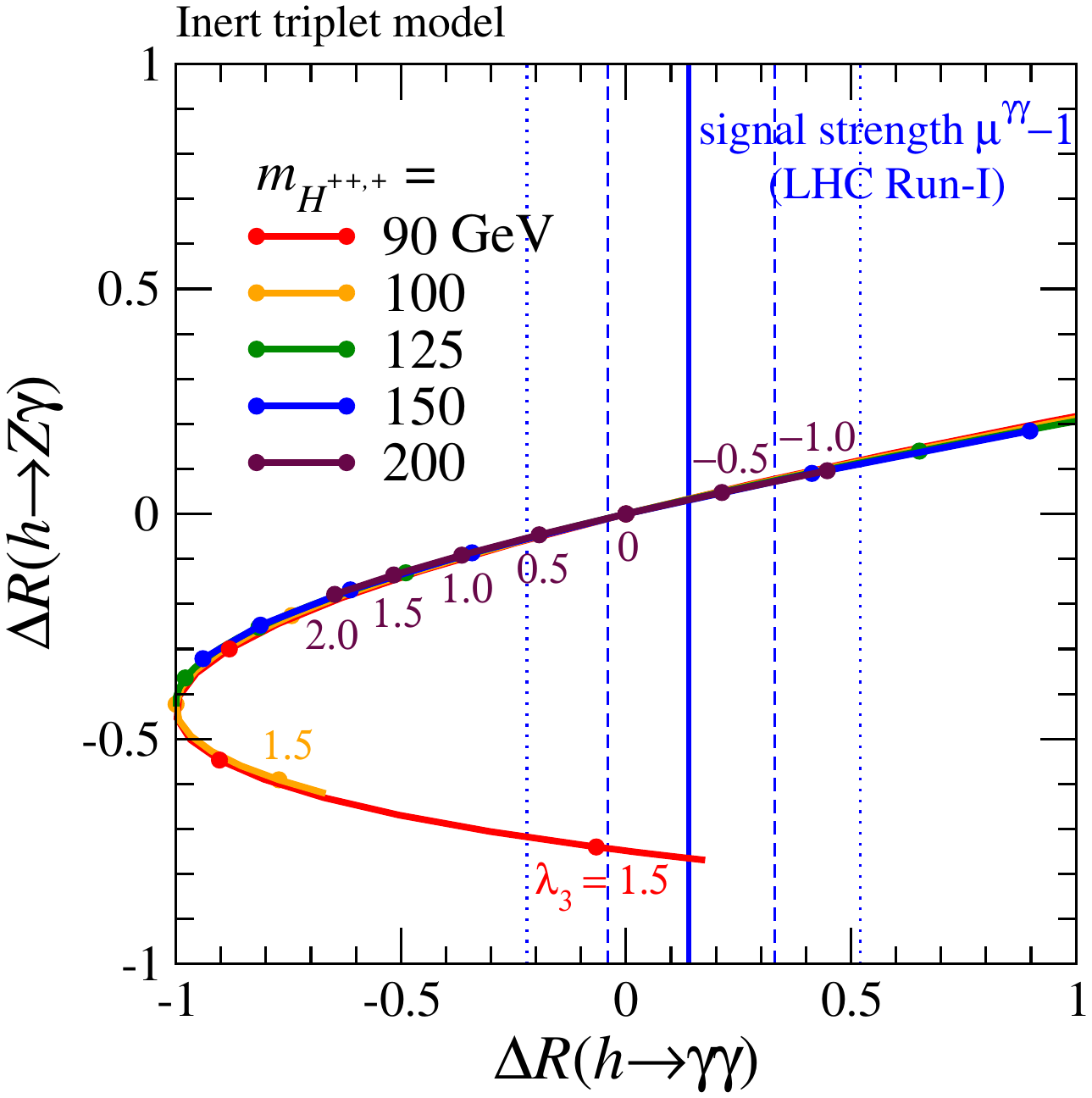}
 \caption{
 Correlations between $\Delta R(\haa)$ and $\Delta R(\hza)$ in the IDM (left) and in the ITM (right).
 }
\label{fig:dRaa_za}
\end{figure}

In Fig.~\ref{fig:rs250_thdm}(right), we present correlations between $\draa$ and $\drha$, 
where we still keep $\tan\beta=2$ as an illustration. 
Although we vary $\cba$ as $0<\cba<0.3$, this mixing effect is small, and hence the correlations look similar to ones in the IDM in Fig.~\ref{fig:rs250_idm}(right), except for no allowed parameter region for positive $\draa$ and $\drha$.
In short, the production rate in the THDM tends to reduce from the SM prediction after imposing the theoretical constraints,
and the magnitude of the deviation is minor after imposing all the current experimental constraints. 

Before turning to the case for $\sqrt{s}=500$~GeV, we mention the $\hza$ decay,
which is closely related to the $\haa$ decay. 
As seen in Fig.~\ref{fig:dRaa_za}, $\draa$ and $\drza$ are strongly correlated in the IDM and the ITM.
The correlation in the THDM is very similar to that in the IDM, and hence we omit it. 
The sensitivity to the charged scalar loops for the $\hza$ decay is much 
weaker than for the $\haa$ decay.
Moreover, from the experimental point of view, the observation of the $\hza$ decay is much more difficult than the $\haa$ decay; 
the current LHC data only give the upper limit on the production cross section times the branching ratio for $pp\to h\to Z\gamma$ as about seven times the SM prediction~\cite{Aaboud:2017uhw,Sirunyan:2018tbk}. 
An interesting thing in the ITM is that, 
for the case of $m_{H^{+}}=m_{H^{++}}=90$~GeV with $\lam_3=1.5$, leading to the large enhancement for the $h\gamma$ production,
we can retain the $\haa$ decay as in the SM, but the $\hza$ decay should be suppressed 
by about 70\% from the SM prediction. 
In other words, if we observe the $\hza$ decay as in the SM in future experiments, we could not expect such $h\gamma$ enhancement.

\subsection{Results at $\sqrt{s}=500$~GeV}

As we have seen in Fig.~\ref{fig:xsec_sm}, the $\eeha$ process in the SM has the second peak around $\sqrt{s}=500$~GeV, 
the cross section of which is about a half of that at $\sqrt{s}=250$~GeV.%
\footnote{The $K$ factor of the next-to-leading-order QCD corrections is about $-0.4$\% and $14$\% 
at $\sqrt{s}=250$ and 500~GeV, respectively~\cite{Sang:2017vph}.}
Here, we repeat the exact same analyses in the previous subsection, but consider the collision energy at $500$~GeV,
where one can find rather different parameter dependences of the cross sections and the $\draa$--$\drha$ 
correlations below.

\begin{figure}
 \includegraphics[height=0.4\textheight]{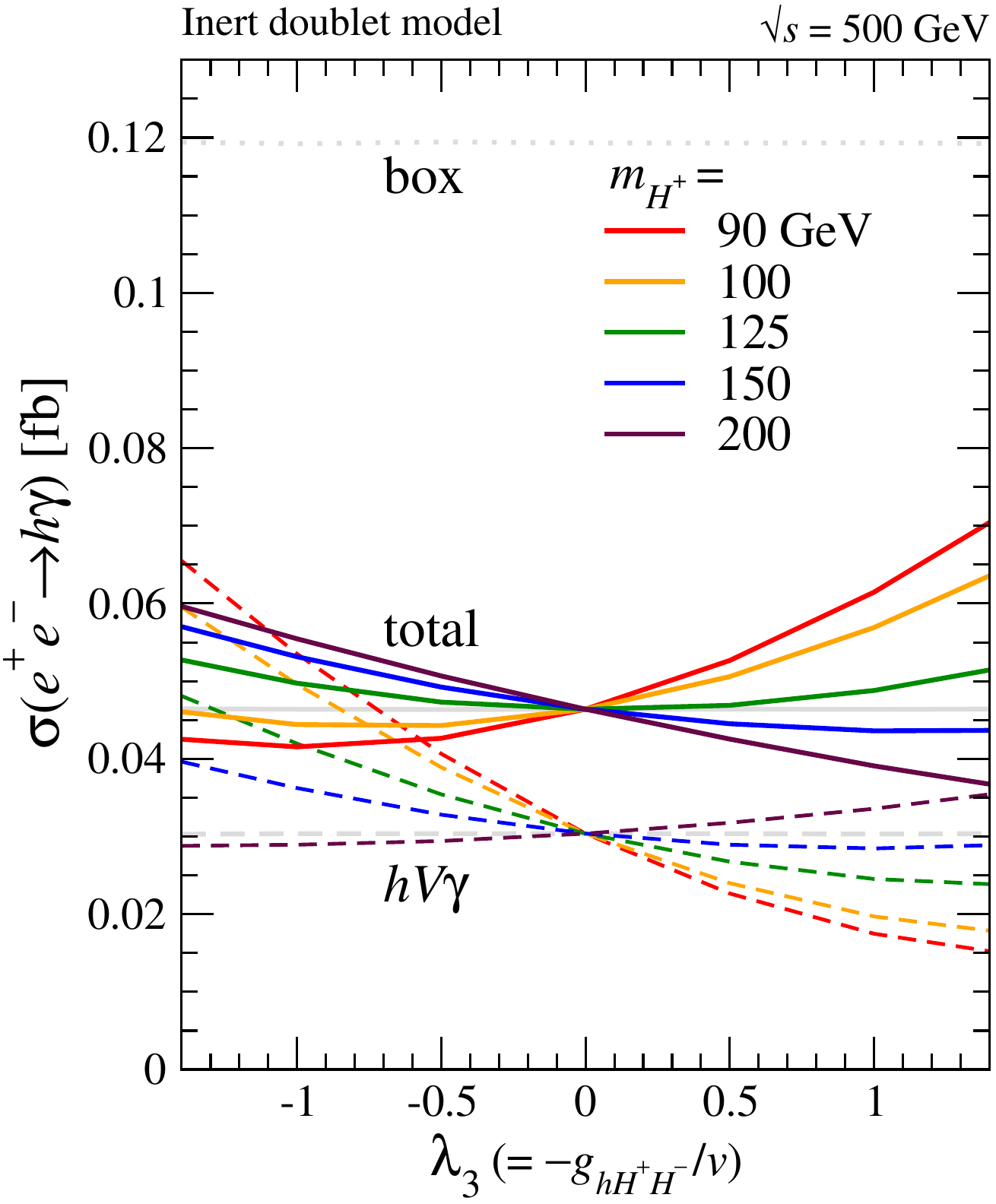}\qquad
 \includegraphics[height=0.4\textheight]{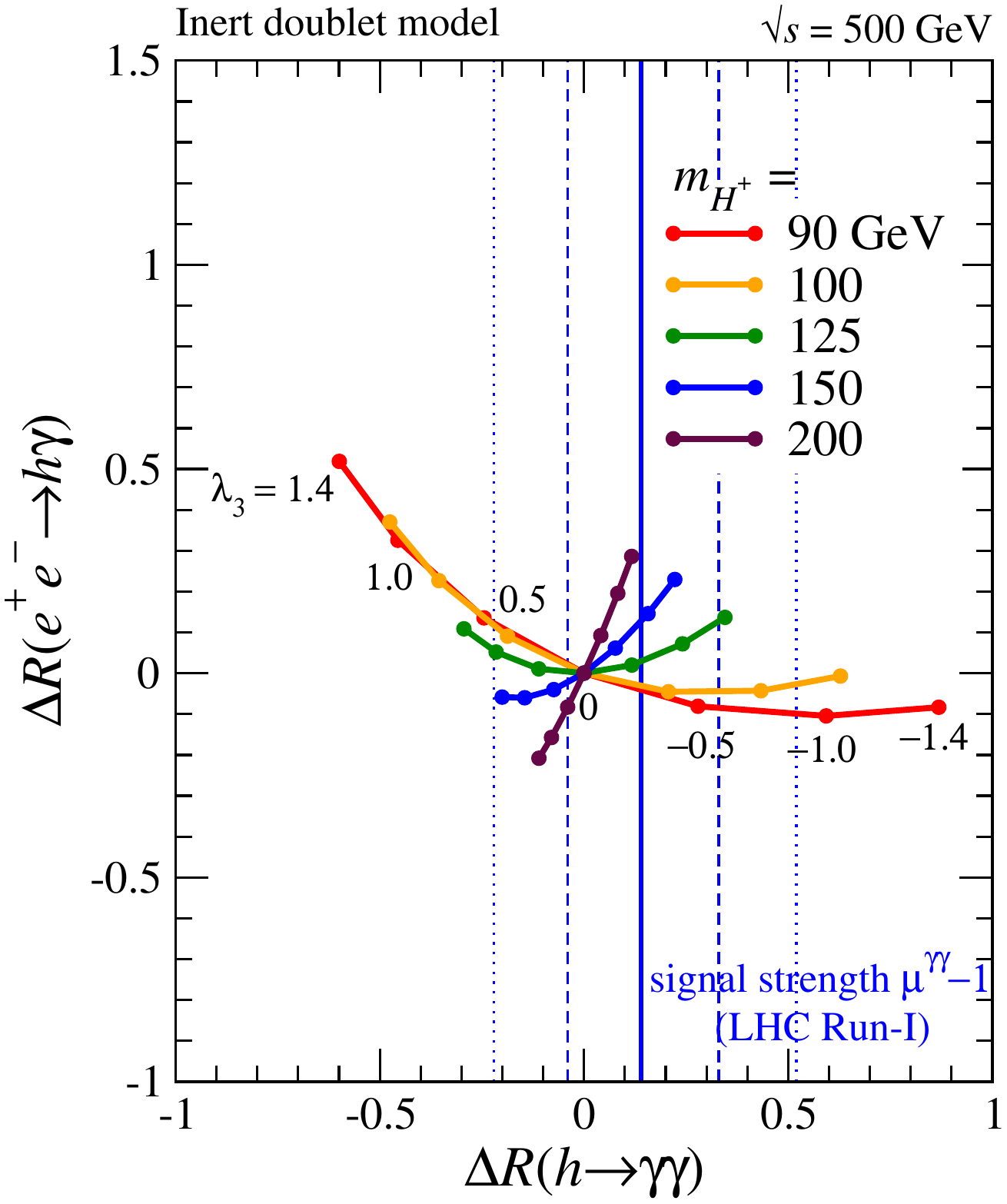}
 \caption{
 Same as Fig.~\ref{fig:rs250_idm}, but at $\sqrt{s}=500$~GeV.
 Total cross sections for $\eeha$ as a function of $\lam_3$ (left) and 
 correlations between $\Delta R(\haa)$ and $\Delta R(\eeha)$ (right) in the IDM. 
 }
\label{fig:rs500_idm}
\end{figure}

\begin{figure}
 \includegraphics[height=0.4\textheight]{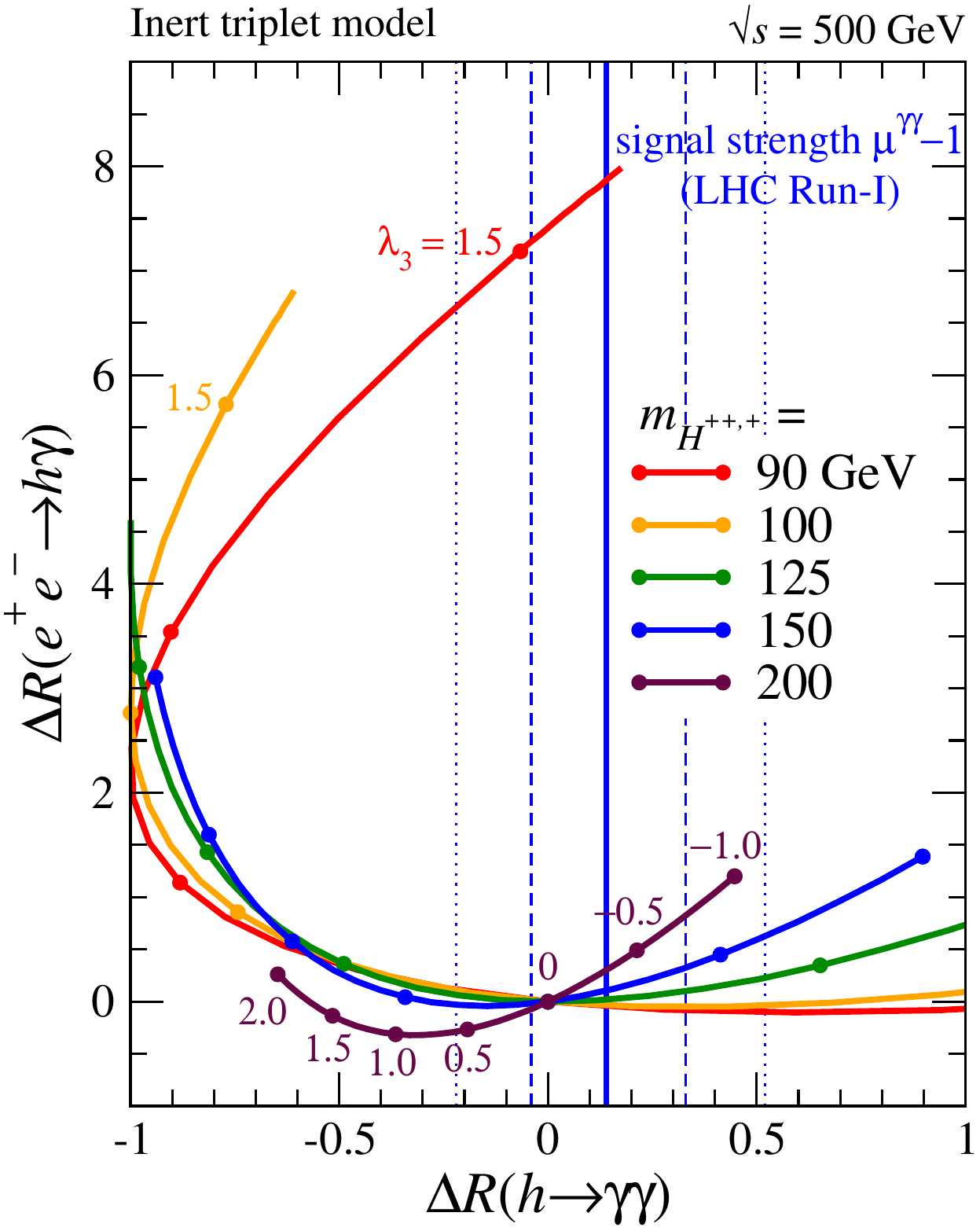}\qquad
 \includegraphics[height=0.4\textheight]{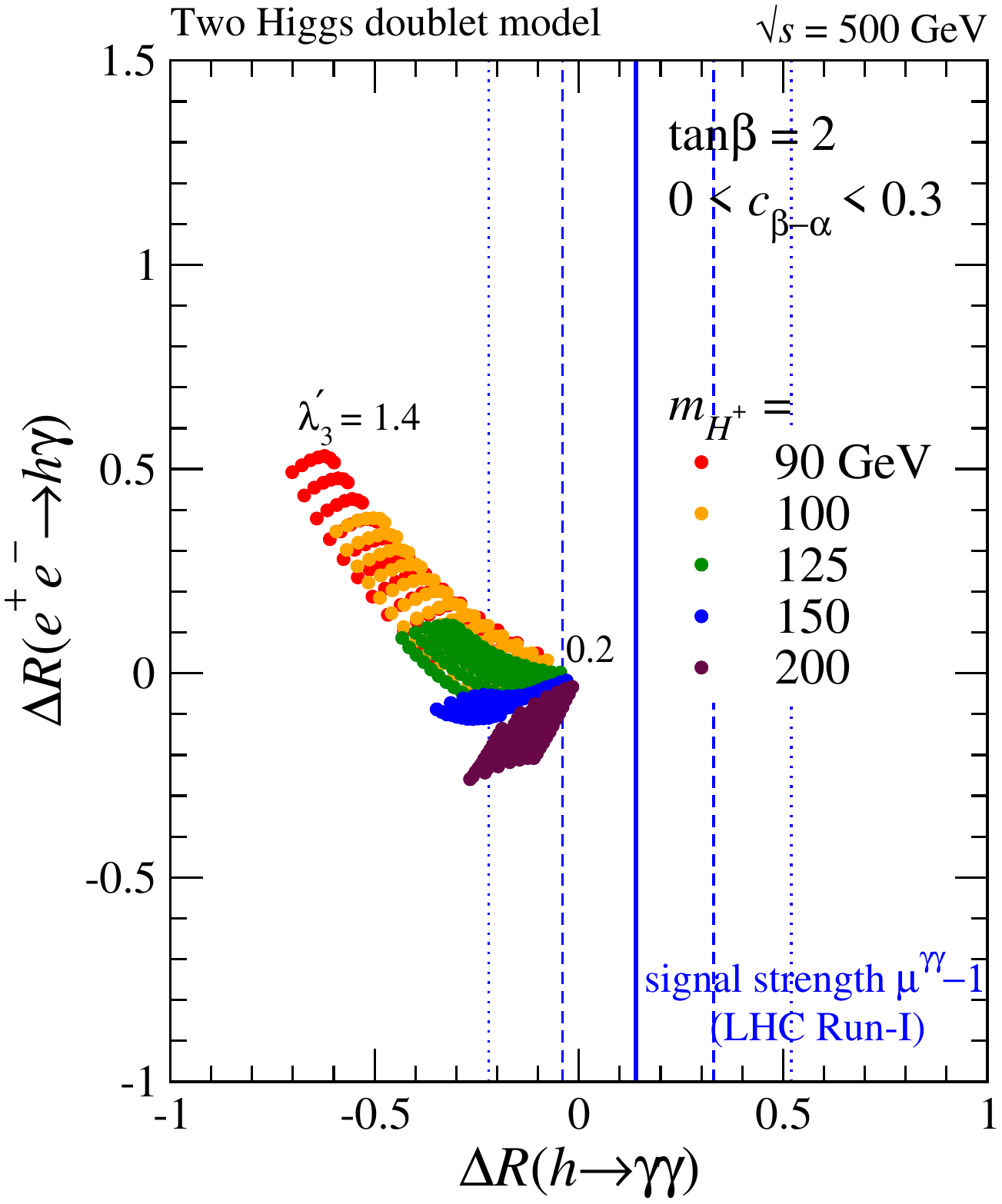}
 \caption{
 Correlations between $\Delta R(\haa)$ and $\Delta R(\eeha)$ in the ITM (left) and in the THDM (right)
 at $\sqrt{s}=500$~GeV.
 }
\label{fig:rs500}
\end{figure}

In Fig~\ref{fig:rs500_idm}, similar to Fig~\ref{fig:rs250_idm}, we show $\sigma(\eeha)$ as a function of $\lam_3$ (left panel) and $\draa$--$\drha$ (right panel) in the IDM at $\sqrt{s}=500$~GeV.
Unlike the case at $\sqrt{s}=250$~GeV, 
the production rate can be enhanced for 
$\mc\geq125$~GeV with negative $\lam_3$ as well as for  
$\mc\leq125$~GeV with positive $\lam_3$. 
Although the $\lam_3$ dependence of the cross sections from the $hV\gamma$ diagrams (dashed lines)
is similar to the case at $\sqrt{s}=250$~GeV;
i.e., the $hV\gamma$ contributions are enhanced (reduced) for negative (positive) $\lam_3$ for light charged scalars,
the contributions of each amplitude are rather different from the $\sqrt{s}=250$~GeV case, resulting in the different $\lam_3$ dependence of the total cross sections (solid lines).
At $\sqrt{s}=500$~GeV, the box contributions (a gray dotted line) are dominant (see also Fig.~\ref{fig:xsec_sm}), 
and the top-loop contributions are also significant. 
Moreover, all the particles in the loops, including charged scalars of our benchmark, can be on shell; i.e.,
all the amplitudes become complex, leading to intricate interference effects.
Since the Higgs-boson decay does not depend on $\sqrt{s}$, we find rather different correlations between $\draa$ and $\drha$ for different masses of the charged scalars.
Especially, the correlations at $\sqrt{s}=250$ and 500~GeV are opposite for $\mc=90-100$~GeV. 

Figure~\ref{fig:rs500} presents correlations $\Delta R(\haa)$ and $\Delta R(\eeha)$ in the ITM (left panel) and in the THDM (right panel)
 at $\sqrt{s}=500$~GeV. 
All the behaviors can be understood from the discussions above. 
We stress that in the ITM we can find a particular parameter region where the $h\gamma$ production is enhanced by a factor of $7.5-9$, but the $\haa$ decay still remains as in the SM, if the mass of the charged Higgs boson is
around 90~GeV. 
Similarly to the $\sqrt{s}=250$~GeV case, 
such parameter region predicts the suppression of the $\hza$ decay.

\section{Summary}\label{sec:summary}

We studied Higgs production associated with a photon at $e^+e^-$ colliders
in the IDM, the ITM and the THDM as distinctive new physics benchmark scenarios.
The cross section in the SM is maximal around $\sqrt{s}=250$~GeV,
and we presented how and how much the new physics can enhance or reduce the production rate.
We also discussed the correlations with the $\haa$ and $\hza$ decay rates.
We found that, with a sizable coupling to a SM-like Higgs boson, charged scalars can give considerable contributions to both the production and the decay
if their masses are around 100~GeV.
Under the theoretical constraints from vacuum stability and perturbative unitarity
as well as the current constraints from the Higgs measurements at the LHC,
the production rate can be enhanced from the SM prediction at most by a factor of two in the IDM. 
In the ITM, in addition, we found a particular parameter region, $\mcc=\mc\sim90$~GeV with $\lam_3\sim1.5$, 
where the $h\gamma$ production significantly increases by a factor of about six to eight, 
but the $\haa$ decay still remains as in the SM.
For such a parameter region, on the other hand, the $\hza$ decay is suppressed. 
In the THDM, possible deviations from the SM prediction are minor in the viable parameter space.
We also showed that the dependence of the cross section on the model parameters is rather different 
at $\sqrt{s}=500$~GeV, leading to the possibility of accessing more information on the Higgs sector.

As a final remark, we would like to point out that new physics can affect not only the total cross section but also the differential distribution, 
which would be important for realistic simulation of the signal events to compare with the background. 

\section*{Acknowledgements}
The work of S.K. is supported in part
by Grant-in-Aid for Scientific Research on Innovative Areas, the Ministry of Education, Culture, Sports, Science and Technology, No.~16H06492 and No.~18H04587, 
and Grant H2020-MSCA-RISE-2014 No.~645722 (Non-Minimal Higgs).
The work of K.M. was supported in part by JSPS KAKENHI Grant No.~18K03648.
The work of K.S. was supported in part by JSPS KAKENHI Grant No.~18J12866.

\begin{appendix}
\section{$h\gamma\gamma$ and $hZ\gamma$ vertex functions}\label{sec:app}

In this Appendix we present analytical expressions for the $h\gamma\gamma$ and $hZ\gamma$ vertex functions
at one-loop level in extended Higgs models.

The $h\gamma\gamma$ vertex is loop induced and does not need to be renormalized, so the vertex functions $\Gamma_{h\gamma\gamma}^{i}$ ($i=1,2$) directly correspond to its  1-particle irreducible (1PI) diagram contributions
\begin{align}
 \Gamma_{h\gamma\gamma}^{ i}(p_1^2,p_2^2,p^2_h)=\Gamma_{h\gamma\gamma}^{i, {\rm 1PI}}(p_1^2,p_2^2,p^2_h).
\end{align}
The 1PI diagram contributions are divided into fermion loop, $W$-boson loop and charged scalar loop contributions, which are expressed in terms of Passarino--Veltman functions~\cite{Passarino:1978jh}.%
\footnote {For convenience,  we abbreviate $C$ and $B$ functions as 
$C_i(p_1^2,p_2^2,p^2_h;m_X,m_Y,m_Z)\equiv C_i(X,Y,Z)$ and
$B_i(p^2;m_X,m_Y)\equiv B_i(p^2;X,Y)$, respectively.}
Regardless of a choice of models, each contribution is expressed as
\begin{align}
 \Gamma&_{h\gamma\gamma}^{\rm 1, 1PI}(p_1^2,p_2^2,p^2_h)_F
  =-\kappa_f\frac{e^2gm_W}{16\pi^2}\sum_f4N_cQ_f^2\frac{m_f^2}{m_W^2}
   \Big\{\frac{1}{2}(p^2_1+p^2_2-p^2_h)C_0+4C_3\Big\}(f,f,f), \\ 
 \Gamma&_{h\gamma\gamma}^{\rm 2, 1PI}(p_1^2,p_2^2,p^2_h)_F
  =-\kappa_f\frac{e^2gm_W}{16\pi^2}\sum_f4N_cQ_f^2\frac{m_f^2}{m_W^2}(C_0+4C_2)(f,f,f), \\ 
 \Gamma&_{h\gamma\gamma}^{\rm 1, 1PI}(p_1^2,p_2^2,p^2_h)_V 
  =\kappa_V\frac{e^2gm_W}{16\pi^2}\Big\{
  (5p_1^2+5p_2^2-7p^2_h-m_h^2)C_0+4\Big(6+\frac{m_h^2}{m_W^2}\Big)C_3
  \Big\}(W,W,W), \\  
 \Gamma&_{h\gamma\gamma}^{\rm 2, 1PI}(p_1^2,p_2^2,p^2_h)_V 
  =\kappa_V\frac{4e^2gm_W}{16\pi^2}\Big\{4C_0+
   \Big(6+\frac{m_h^2}{m_W^2}\Big)C_2\Big\}(W,W,W), \\
 \Gamma&_{h\gamma\gamma}^{\rm 1, 1PI}(p_1^2,p_2^2,p^2_h)_S 
  =-\frac{8e^2}{16\pi^2}\sum_S g_{hSS^\ast} Q_S^2 C_{3}(S,S,S), \label{Ghaa1} \\ 
 \Gamma&_{h\gamma\gamma}^{\rm 2, 1PI}(p_1^2,p_2^2,p^2_h)_S 
  =-\frac{8e^2}{16\pi^2}\sum_S g_{hSS^\ast} Q_S^2 C_2(S,S,S), \label{Ghaa2}
\end{align}
where $C_2$ and $C_3$ are defined by
\begin{align}
 C_2(X,X,X)&=C_{12}(X,X,X)+C_{23}(X,X,X), \\
 C_3(X,X,X)&=C_{24}(X,X,X)-\frac{1}{4}B_0(p^2_h;X,X).
\end{align}
The scaling factors $\kappa_{f,V}$ and the trilinear couplings between the SM-like Higgs boson and charged scalars $g_{hSS^*}$ depend on each extended Higgs model.
See Eqs.~\eqref{kappa_idm} and \eqref{ghHH_idm} in the IDM,  
Eqs.~\eqref{kappa_itm}, \eqref{ghHH2_itm} and \eqref{ghHH_itm} in the ITM, and
Eqs.~\eqref{kappa_thdm} and \eqref{ghHH_thdm} in the THDM.

The $hZ\gamma$ vertex is also loop induced and needs counterterms to remove ultraviolet divergence unlike the $h\gamma\gamma$ vertex. 
Hence, the renormalized $hZ\gamma$ vertex functions $\Gamma_{hZ\gamma}^{i}$ ($i=1,2$) are written as
\begin{align}
 \Gamma_{hZ\gamma}^{i}(p_1^2,p_2^2,p^2_h)=\Gamma_{hZ\gamma}^{i, {\rm 1PI}}(p_1^2,p_2^2,p^2_h)+\delta\Gamma^{i}_{hZ\gamma}.
\end{align}
The 1PI diagram contributions from fermion, $W$-boson and charged scalar loops are expressed, respectively, as
\begin{align}
 \Gamma_{hZ\gamma}^{\rm 1, 1PI}(p_1^2,p_2^2,p^2_h)_F 
  &=-\kappa_f\frac{egg_Zm_W}{16\pi^2}\sum_f   
  4N_cQ_fv_f\frac{m_f^2}{m_W^2}\Big\{\frac{1}{2}(p^2_1+p^2_2-p^2_h)C_0+4C_3\Big\}(f,f,f), \\ 
 \Gamma_{hZ\gamma}^{\rm 2, 1PI}(p_1^2,p_2^2,p^2_h)_F 
  &=-\kappa_f\frac{egg_Zm_W}{16\pi^2}\sum_f4N_cQ_fv_f\frac{m_f^2}{m_W^2}(C_0+4C_2)(f,f,f),\\  
 \Gamma_{hZ\gamma}^{\rm 1, 1PI}(p_1^2,p_2^2,p^2_h)_V 
  &= \kappa_V\frac{egg_Zm_W}{16\pi^2}\Big[
  2B_0(p_2^2;W,W) \notag\\ \notag
  &\quad+\{(5p^2_1+5p_2^2-7p^2_h)c_W^2-2p^2_1+p^2_2+p^2_h 
   +3m_W^2+m_h^2s_W^2\}C_0(W,W,W)  \\ 
  &\quad+2\Big\{12c_W^2-2+\frac{m_h^2}{m_W^2}(2c_W^2-1)\Big\}C_3(W,W,W)\Big], \\ 
 \Gamma_{hZ\gamma}^{\rm 2, 1PI}(p_1^2,p_2^2,p^2_h)_V 
  &=\kappa_V\frac{egg_Zm_W}{16\pi^2}\Big[4(4c_W^2-1)C_0 
  +2\Big\{12c_W^2-2+\frac{m_h^2}{m_W^2}(2c_W^2-1)\Big\}C_2 \Big] (W,W,W),\\
 \Gamma_{hZ\gamma}^{\rm 1, 1PI}(p_1^2,p_2^2,p^2_h)_S
  &=-\frac{8eg_Z}{16\pi^2}\sum_Sg_{hSS^\ast}Q_S(I_3^S-Q_Ss_W^2)C_{3}(S,S,S), \\ 
 \Gamma_{hZ\gamma}^{\rm 2, 1PI}(p_1^2,p_2^2,p^2_h)_S
  &=-\frac{8eg_Z}{16\pi^2}\sum_Sg_{hSS^\ast}Q_S(I_3^S-Q_Ss_W^2)C_2(S,S,S), 
\end{align}
where $v_f=I_f/2-Q_fs_W^2$ and $g_Z=\sqrt{g^2+g^{\prime 2}}$.
$Q_S$ and $I_3^S$ denote electric charge and isospin for charged scalars, respectively.  
The counterterm contributions in an improved on-shell scheme~\cite{Kanemura:2017wtm}, which is applied in {\sc H-COUP}, are written by
\begin{align}
 \delta \Gamma_{hZ\gamma}^{\rm 1}=
 -\kappa_V\frac{egg_Zm_W}{16\pi^2}2B_{0}(0;W,W), \quad \delta \Gamma_{hZ\gamma}^{\rm 2}=0.
\end{align}
We note that extra Higgs loop contributions do not appear in the counterterms.
Similar to the $h\gamma\gamma$ vertex functions, 
$\kappa_{f}$, $\kappa_V$ and $g_{hSS^\ast}$ are given in each model.
   
\begin{figure}
 \includegraphics[width=0.5\textwidth]{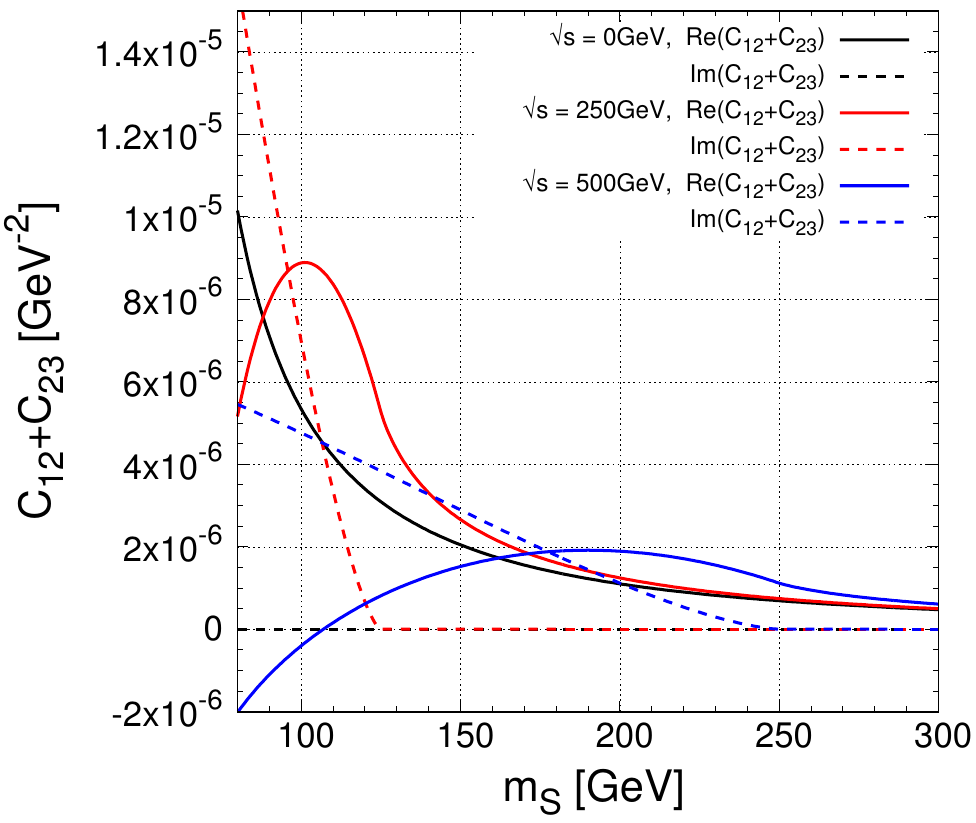}
 \caption{Real and imaginary parts of the loop function $(C_{12}+C_{23})(p_1^2,p_2^2,p_h^2;m_S,m_S,m_S)$ as a function of the mass of the scalar particle $m_S$. External momenta are fixed at $p_1^2=0$, $p_2^2=({0,250,500}~{\rm GeV})^2$, and $p_h^2=m_h^2$.}
\label{fig:loopfunc}
\end{figure}
       
For better understanding of our results in the main text, we look at 
the $m_S$ dependence of $\Gamma^2_{h\gamma\gamma,S}$ for the $\eeha$ production and the $\haa$ decay.
As discussed in Sec.~\ref{sec:hcoup}, $\Gamma^1_{h\gamma\gamma,S}$ is related to $\Gamma^2_{h\gamma\gamma,S}$
due to the gauge invariance.
For the kinematical configuration $p_1^2=0$, $p_2^2=s$, and $p_h^2=m_h^2$, 
$C_2$ can be expressed by only the $C_0$ and $B_0$ functions as
\begin{align}
 C_2(0,s,m^2_h;m_S,m_S,m_S)
  &=\frac{1}{2(s-m^2_h)}\Big[1+2m_S^2C_0(0,s,m^2_h;m_S,m_S,m_S) \notag\\ 
  &\quad+\frac{s}{s-m^2_h}\{B_0(s;m_S,m_S)-B_0(m_h^2;m_S,m_S)\}\Big].
\end{align}
In Fig.~\ref{fig:loopfunc}, 
we show behaviors of the loop function $C_2\,(\equiv C_{12}+C_{23})$ as a function of the mass of scalar particle $m_S$,
where $\sqrt{s}$ is fixed at 0, 250 and 500~GeV, denoted by black, red and blue lines, respectively.
For $m_S>\sqrt{s}/2$, the loop function becomes pure real, and goes to zero 
in the large mass limit. 

\end{appendix}

\bibliography{bibhcoup}
\bibliographystyle{JHEP}

\end{document}